\documentclass[12pt,twoside]{article}
\usepackage{correl}
\def\TheTitle{Competition between Kondo Effect and RKKY Coupling}
\def\TheHeading{Kondo Effect and RKKY Coupling}
\def\TheAuthor{Stefan Kettemann (
to be published  in
Lecture Notes of the Autumn School on Correlated Electrons 2024,  Correlations and Phase Transitions, edited by 
Eva Pavarini and Erik  Koch)}
\def\TheInstitute{Constructor University Bremen}
\def\TheAddress{Campus Ring 1,  28759 Bremen}
\def\TheChapter{}           

\begin{document}
\MakeTitle
\section{Introduction}
    Localized  magnetic moments \index{magnetic  moments} in metals,
 for example  formed by Fe-atoms in gold, 
      interact with the spins of 
     itinerant electrons  in the Fermi sea 
   via   exchange couplings $J.$ This  results in 
   spin dependent electron scattering,   in addition to  potential scattering  from the 
    impurity potential. Depending on magnetic moment  density 
   $n_{\rm M}$,  magnetic impurity spin $S$,   magnitude and  sign of  
   exchange couplings \index{exchange coupling}  $J$ and  temperature $T,$
    the metal  settles for  one of a diverse set of 
  quantum  phases, each with very different degrees of spin and charge correlations.

\begin{figure}[t!]
 \centering
 \includegraphics[width=0.5\textwidth]{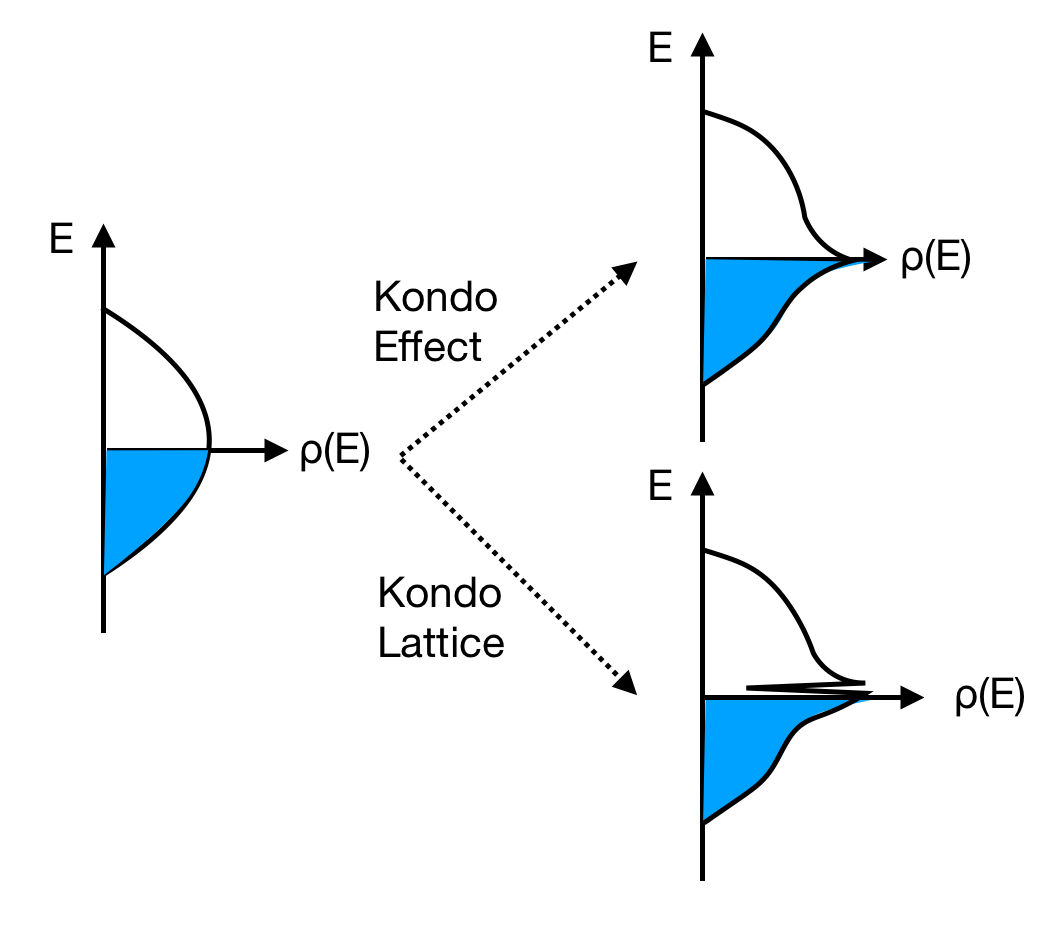}
 \caption{Left: Sketch of typical  density of states $\rho(E) $ of a metal as function of energy E, 
 with states filled up to the Fermi energy $\epsilon_{\rm F}$, as colored in blue.
 Upper Right: For dilute magnetic impurities with spin $S=1/2$, coupled by an antiferromagnetic exchange coupling to the conduction electron spins, a Kondo resonance  \index{Kondo resonance}  of width $k_{\rm B} T_{\rm K }$
   forms. Lower Right: At sufficiently large density 
  a  Kondo lattice forms with  a narrow band at the Fermi energy. }
 \label{ks}
\end{figure}

    If the  exchange coupling $J$  with impurity spin $S=1/2$ is antiferromagnetic,
     all conduction electrons
       compete to form a singlet with this localized spin, if $J$ is not sufficiently strong to bind  and localize one of the  electrons completely
        into a singlet. 
        This competition leads to  strongly enhanced 
        magnetic and normal scattering, measurable as enhanced electrical resistivity,  as the temperature is lowered towards and below a  temperature $T_{\rm K}$. 
         This effect was explained  by Kondo\cite{Kondo} and is now known as 
 the Kondo effect.\index{Kondo effect} 
 The resistance minimum as function of temperature close to  the 
 Kondo temperature $T_{\rm K}$\cite{Hewson} occurs since above $T_{\rm K}$  the resistance   decays  with temperature, 
 as typical for a metal,  while it increases at and   below  $T_{\rm K}$ due to Kondo  enhanced scattering rate.  
   While  such magnetic moments are paramagnetic, contributing  a Curie magnetic susceptibility 
   $\chi \sim 1/T$ at higher temperature $T > T_{\rm K},$ at lower temperature 
    their contribution to the magnetic susceptibility saturates to  $\chi \sim 1/ T_{\rm K}$. 
 More recently, it was found that  mesoscopic metal wires with dilute magnetic impurities
show a  pronounced peak  at  $T_{\rm K}$  in the temperature dependent  dephasing rate,  which governs quantum corrections to the conductance, the so called weak localization corrections\cite{Mohanty,Pierre,Mallet}, allowing high precision  studies of  the Kondo screening. 
The Kondo temperature $T_{\rm K}$ is a functional of the local exchange coupling $J$ and the local density of states $\rho,$ at and in the vicinity of the Fermi energy $\epsilon_F.$
 Remarkably, as the temperature is lowered further, a portion of the conduction electrons settle for a joint screening of the magnetic impurity spin and form the so called  Kondo singlet\index{Kondo singlet}, a  highly correlated  many body state. For the remaining conduction electrons the  magnetic impurity spin seemingly  then disappears. The electrons settle  then
  again to form a Fermi liquid,
 albeit 
 with enhanced density of states at the Fermi energy, forming a narrow resonance peak 
 of width $k_B T_{\rm K}$, as sketched in Fig. \ref{ks} (upper Right).
  As a consequence, the spin scattering from the magnetic impurity decays to zero, as the temperature is lowered further.  The remaining
enhanced  potential scattering from the  Kondo impurities then contributes to the enhanced low temperature resistance.

   When the concentration of magnetic moments
   in a metallic host 
    is  high, and they form a regular lattice, all Kondo impurities can conspire to 
     form a narrow band at the Fermi energy, as sketched in Fig. \ref{ks} (lower Right),
      below a critical temperature  $T_c< T_{\rm  K}$.
      Then,   the itinerant electrons at the Fermi energy 
     are no longer scattered from the potential of the Kondo impurities but move
    through  the lattice formed by the  Kondo impurities as  
     dressed quasiparticles with strongly enhanced mass, 
     accordingly called {\it heavy fermions}\index{heavy fermions}. 
   This transition to a new state of heavy but itinerant  fermions is experimentally seen,
   when the Fermi energy is in that narrow band,
    in a sudden 
   drop of the resistivity below  a critical temperature $T_c$, where the 
    low temperature coherence sets in,  while  at higher temperatures still the typical 
   Kondo enhanced resistivity from individual Kondo impurities is observable. 
   This is observed for example 
 in the intermetallic crystal Ce Cu$_6$, where the Ce$^{3 +}-$ ions form  at high temperature  a dense lattice of magnetic moments  in the metallic copper host, while at low temperatures 
  a transition to a coherent state of heavy electrons occurs with  a sharp drop of resistivity\cite{nozieresfl,Nozieres1998,Coleman2007,Fulde2012}. 

However, localized  magnetic moments in metals  interact with each other. 
Their magnetic dipole interaction is finite, but is  typically exceeded by far by    indirect exchange couplings between them,
the  so called RKKY couplings\index{RKKY coupling},  mediated by  conduction electrons\cite{RK,K,Y}.
In metals,  RKKY coupling  decays slowly, with a power law of distance $R$ between 
 two magnetic moments. The  RKKY coupling is   a functional of 
  exchange couplings $J,$  local density of states at the Fermi energy at the  locations of the magnetic impurities
  $\rho({\bf r},E_F)$ and  distance $R.$
  Since these couplings
   tend to  quench their spins, it may prevent the Kondo screening by the 
    conduction electrons partially, or even completely, depending on 
    the  amplitude of  local  exchange coupling $J,$  
    the distance $R$ and  temperature $T$. 
  
 \begin{figure}[t!]
 \centering
 \includegraphics[width=1.\textwidth]{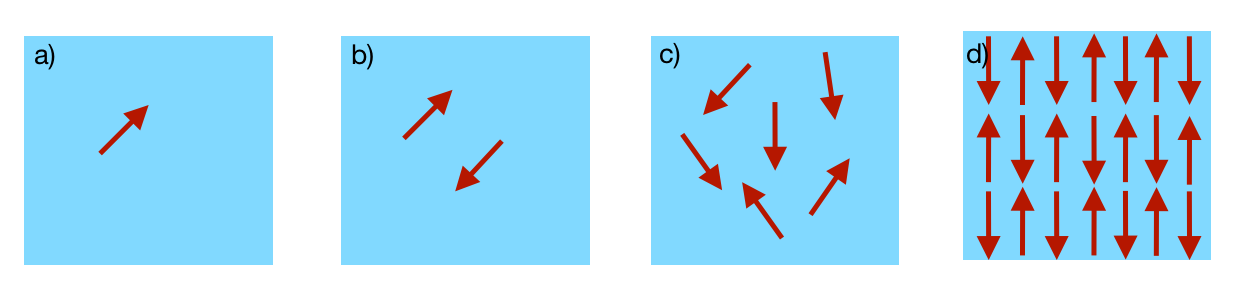}
 \caption{Kondo impurity spins (red) in a metal host (blue): a) Single Kondo impurity, b) Pair of Kondo impurities, c) Dilute Kondo system, d) Kondo lattice.}
 \label{mms}
\end{figure}

 Thus,  there is a   competition between  Kondo screening and   RKKY coupling. 
    Depending on which one wins,  the   system will find itself
    in very  different quantum states.  
     This competition can be studied systematically by increasing the density of magnetic impurities.
     Increasing their concentration, decreases  their average distance $R$ and 
     thereby the typical  RKKY coupling between them increases. 
   In  the very  dilute limit  RKKY couplings can be neglected, and  magnetic impurities can be treated as  a dilute set of single Kondo spins, as shown in Fig. 
   \ref{mms} a).    Increasing their   concentration further, randomly placed magnetic impurities
    may be  modeled by a set of 
    pairs of  magnetic impurities, 
    formed   by those spins which are closest to each other,
    as shown in Fig. 
   \ref{mms} b). 
    At higher concentrations larger clusters of spins, shown in Fig.
   \ref{mms} c) have  to be considered to model their quantum state,  and
    at  still higher concentrations  a connected
    random network of them.  
   When the density of magnetic moments is  high and they  form a regular lattice, as it occurs in  
   f-band materials, a coherent Kondo lattice can form. 
   The competition between  the Kondo effect in this Kondo lattice and the RKKY 
    coupling gives rise to a quantum phase transition between 
    a   heavy fermion
  \index{heavy fermions} state  and an ordered state, 
  which when mapped as function of exchange coupling $J$ is called the Doniach diagram  \index{Doniach diagram} \cite{Doniach77,Kroha17}. 
 
  In these lecture notes we  give an introduction to the theory of this rich  competition between Kondo screening and  RKKY coupling. 
  In section 2 we review the formation of magnetic moments as modeled by the Anderson model. 
  In section 3 we review the theory of the Kondo effect for a single  
  magnetic impurity in a metal host. In section 4
   we  derive the RKKY coupling between  magnetic impurities in a metal host. In section 5 we  review the  Doniach  diagram, and  
   give an introduction to a self consistent
  renormalization group  theory which takes into account both 
   Kondo effect and RKKY coupling 
 between  magnetic impurities, and review results obtained thereby. 
   In section  6 we review the effect of 
    gaps and pseudo gaps on both Kondo effect and  RKKY couplings, and accordingly on their competition. 
  Especially for dilute concentration of magnetic moments, the 
   disorder effects from  randomly distributed impurities  result in
    a distribution of both  Kondo temperatures and  RKKY couplings. Then,  their competition 
     becomes an even more complex problem, as  reviewed in section 7. Moreover, 
      disorder induced  Anderson localization transitions may occur,   which effect both Kondo effect and RKKY coupling severely, and 
      changes their competition, 
      as we review in that section, as well. 
     We conclude with  an outlook and  list of, in our view, most pressing and interesting
      open problems.

     \section{Formation of  magnetic moments}

The first microscopic model for the formation of magnetic moments in metals was formulated by 
P.W. Anderson\cite{anderson}. He showed that local moments 
can form from localized $d$- or $f-$levels which  are weakly coupled by hybridization  to the conduction electrons, when the repulsive 
  Coulomb interaction $U>0$
 between two electrons on these localized levels   
  is sufficiently large. He found furthermore, quite surprisingly,  that the resulting  local moments develop an antiferromagnetic coupling with the  spins of the surrounding electron liquid. 
The formation of magnetic moments is described by 
the Anderson model,\index{Anderson model} where  a  $d$- or $f$-level is weakly 
coupled to  conduction band electrons, as modelled  
 by the Hamiltonian\cite{anderson}
\begin{eqnarray}
\label{Anderson}
H &=& \sum_{n,\sigma} E_{n \sigma} \hat{n}_{n \sigma} +
 \sum_{\sigma}  \varepsilon_{d \sigma} \hat{n}_{d \sigma} + U \hat{n}_{d +} \hat{n}_{d -}
 \nonumber \\ &+& \sum_{n,\sigma} \left( t_{n d} c^+_{n \sigma} d_{
 \sigma } + t_{d n} d^+_{ \sigma} c_{n \sigma } \right),
\end{eqnarray}
where electrons in   a conduction band state 
$|n\rangle$  with eigen energy $E_{n \sigma}$   are annihilated and created by fermion operators 
$c_{n \sigma },c^+_{n \sigma }$ 
with spin index  $\sigma =+,-.$  The corresponding 
density operator is
  $\hat{n}_{n \sigma} = c^+_{n \sigma } c_{n \sigma }$. In the following, we assume spin degeneracy of the conduction band states
  $E_{n \sigma} =E_n.$
  The annihilation and creation operators of electrons in the $d$-level are $d_{ \sigma},d^+_{ \sigma}$
with  density operator $ \hat{n}_{d \sigma}  = d^+_{ \sigma} d_{ \sigma}.$
   The  $d$-level can either be  in a magnetic state, when it is occupied by a single electron with energy 
$\varepsilon_{d \sigma}$,
which  can be in one of  two spin states $\sigma =+,-$. We assume that these two states   form a Kramer's doublet, with  energy 
$\varepsilon_{d},$ degenerate in the spin $\sigma$.
Furthermore, it can be in a nonmagnetic  state when  unoccupied or  
  when doubly occupied, with vanishing total spin and 
total energy  
$2 \varepsilon_d + U$.  
  In order that the ground state  is magnetic, the energy of the singly occupied states must be lower than 
   the energy of the  unoccupied state, as well as   the one of the  doubly occupied state, 
   requiring $\varepsilon_d <0$ and $\varepsilon_d + U  > 0$. Thus,  the repulsion must be stronger than the bound state energy of a single electron,
     $U > - \varepsilon_d $. At finite temperature $T$,  the $d$-level   remains magnetic 
      as long as $T$ is lower than  the energy cost for such valence fluctuations, 
      $ T< {\rm  min} ( \varepsilon_d + U,  - \varepsilon_d)$. 
     However, the hybridization 
between the $d$-level and the conduction band state 
$|n\rangle$, as given  by  the matrix elements 
$t_{d n} = t_{n d}^*,$ may change this  ground state. 
  To study its effect  on the magnetic states, one can 
project  nonmagnetic higher energy states, where the
$d$-level is doubly occupied or unoccupied, out of the Hilbert space of the $d$-level. 
 This was done by Schrieffer and Wolff\cite{Schrieffer1966}, who thereby showed that 
     the spin on the $d$-level is coupled by an antiferromagnetic  exchange interaction $J$ with the 
      spins of the conduction electrons. 
      Performing this  Schrieffer-Wolff transformation 
one obtains the  Kondo Hamiltonian \index{Kondo Hamiltonian} in its most general form
\begin{equation} \label{KH}
H_K  = \sum_{n,\sigma} E_n \hat{n}_{n \sigma} +   \sum_{n,n'} J_{n n'}  \left[ S^+ c^+_{n +}  c_{n' -}  +  S^- c^+_{n -}  c_{n' +}  +  S_z
( c^+_{n +}  c_{n' +} - c^+_{n -}  c_{n' -})
\right]
\end{equation}
where   ${\bf S}$ is the spin   vector  operator of the localized moment,
written here in terms of  $ S^+ = S_x + \i S_y$ and  $ S^- = S_x - \i S_y,$
 and its $z$-component $S_z$. 
  The matrix elements of the exchange coupling \index{exchange coupling} in the basis of
the conduction electron  eigen states $|n\rangle$  are  found to be given by 
\begin{equation}
J_{n n'} =  t_{n d}\, t_{d n'} \left( \frac{1}{U+  \varepsilon_d - E_{n'} } +  \frac{1}{- \varepsilon_d + E_{n} }   \right) .
\end{equation}
The hopping matrix element connecting  the localized $d$-state
$\phi_d({\bf r})$ to the conduction band state $\psi_n({\bf r})$ is
 related to the atomic potential  $\hat{V}$  by the hybridization integral
\begin{equation}
t_{d n} = \langle d | \hat{V} |n \rangle =    \int d^dr\, \phi_d^\ast({\bf r})\, V({\bf r})\,
\psi_n({\bf r}). 
\end{equation}
 For
an impurity state strongly localized on a length scale $a_0$ at position  ${\bf r}$
in a $d$-dimensional sample,  one  can simplify that 
expression with  the hybridization parameter  $t$ to 
\begin{equation}
t_{d n} \approx  t  a_0^{d}  \phi^\ast_d({\bf r})\, \psi_n({\bf r}).
\end{equation}
  Assuming that  both nonmagnetic states have the same energy
    $U/2$,  one arrives at  the symmetric Kondo model.
In this approximation, the Kondo Hamiltonian can be written with the superexchange term  in the form of  a Heisenberg Hamiltonian 
\cite{Kondo,Coleman2007}, 
\begin{equation} \label{HHM}
H^0_K = \sum_{n,\sigma} E_n \hat{n}_{n \sigma} +  J   \vec{{\bf S} } \vec{{\bf s}} ({\bf r}),
\end{equation}
 with $J = 4 t^2/U >0.$ Thus,  the superexchange  interaction  \index{superexchange interaction}  is indeed 
antiferromagnetic \index{antiferromagnetic} . 
The matrix elements of the conduction band spin density vector  operator $\vec{{\bf s}} ({\bf r})$ at the site of the $d$-level, ${\bf r}$
 are   given by 
\begin{equation} \label{sigma} 
\vec{ {\bf s}}_{\alpha \beta } ({\bf r})  = 
\sum_{n,n'}  \psi^*_{n'}({\bf r})  \psi_n({\bf r} ) c^+_{n \alpha}  c_{n' \beta} \vec{ \sigma}_{\alpha \beta },
\end{equation}
where  $\vec{ \sigma}$ is  the  vector of Pauli matrices $\vec{ \sigma} = (\sigma_x, \sigma_y, \sigma_z)$. 
Here,  we used $a_0^d |\phi_d({\bf r})|^2 =1,$ since the  intensity  $|\phi_d({\bf r})|^2$ in the d-level 
 is localized in the volume  $a_0^d$.

    \section{Kondo efffect: screening of magnetic moments}

When the  bare  antiferromagnetic exchange interaction $J$ is too weak to bind a single conduction electron into 
 a singlet state, all conduction electrons in the vicinity of the Fermi energy become excited by   scattering  from the magnetic impurity spin.
 Integrating out all  these excitations, of conduction electrons to energy levels $E_m$ 
above the Fermi energy and  of  hole excitations below the Fermi energy, 
  one finds that the 
 exchange 
   interaction $J$ becomes thereby enhanced. 
   Performing perturbation theory 
to
second order in $J$, there are two processes to be considered: (i) The
scattering due to the exchange coupling $J$ of an electron from initial state
$|n\rangle$ to a state $|l\rangle$ at the Fermi energy via an
intermediate state $|m\rangle$, which can be of either spin. This process is proportional to the
probability that state $|m \rangle$ is not occupied, $1-f(E_m)$, where
$f(E)$ is the Fermi distribution function. (ii) The reverse process, in which a
hole is scattered from the state $|l\rangle$ to the state $|n \rangle$
via the occupied  state $|m\rangle$ which is accordingly proportional
 to 
 the occupation factor $f(E_m)$. Thereby, one  finds that the Kondo exchange Hamiltonian acquires an additional 
  term so that the total exchange coupling 
coupling becomes 
\begin{equation}
\label{eq:poorman}
\tilde{J}_{nl} = J_{nl} \left[ 1 + \frac{J}{2 N} \sum_{m,\sigma} \frac{
Vol. | \psi_m( {\bf r}) |^2 }{E_m - E_F}
 \tanh \left( \frac{E_m - E_F}{2 T} \right) \right],
\end{equation}
where $Vol. = L^d$ is the volume of the electron system. For positive exchange coupling, $J>0$, 
the correction term is positive as well. Moreover, 
this perturbation theory  
diverges as
the temperature is lowered.  
 Defining the Kondo temperature  \index{Kondo temperature} as the
temperature where perturbation theory breaks down since  the second-order correction to the exchange coupling
becomes equal to the bare coupling, we find   in this   1-loop approximation \index{1-loop approximation} , 
 that  the Kondo
temperature at site ${\bf r}$ of a spin-$1/2$-impurity is determined by the equation \cite{Nagaoka65,Suhl65}
\begin{equation}
\label{eq:FTK}
1= \frac{J}{2 N} \sum_{n,\sigma}  \frac{L^d |\psi_n ({\bf r})|^2}{E_n-E_F} \tanh 
\left(  
\frac{E_n-E_F}{ 2 T_{\rm K} ( {\bf r} )} \right),
\end{equation}
with $N$ the total number of energy levels, including spin degeneracy,  in a finite sample of linear size $L$ and dimension $d$.
 $|\psi_n ({\bf r})|^2$ is the probability density of the eigen state
 at  site ${\bf r}$. 
 
An equivalent expression can be derived from a renormalization group \index{renormalization group} analysis.
Integrating successively high energy excited states
 at  energy scale $\Lambda$ above and below the Fermi energy yields 
   a renormalized coupling  $\tilde{J} (\Lambda)$, governed by the RG flow equation\cite{AndersonYuval,Zarand96}. 
For 
 a magnetic moment at site ${\bf r}$ with exchange coupling
   $J$ and   local  density of states at energy $\epsilon$,
   $\rho({\bf r},\epsilon) $,
 the renormalization of the effective coupling $\tilde{J}(\Lambda)$
 at energy scale $\Lambda$, above and below the Fermi energy, is
    found    in 1-loop approximation  to be given by
\begin{eqnarray}  \frac{d \tilde{J}}{d \ln \Lambda} &=& - \tilde{J}^{2} \frac{V_a}{2} (\rho({\bf r},\epsilon_F +\Lambda)  +
 \rho({\bf r},\epsilon_F -\Lambda)),
\label{rg}
\end{eqnarray} 
 where $V_a = L^d/N$  is the atomic volume, which is often set equal to one,  we will keep it  for clarity.
  The  solution  of Eq. (\ref{rg})  diverges for small energy scales  $\Lambda \rightarrow 0$.
   Defining here   the Kondo temperature by  the RG scale $\Lambda_K   = k_B T_K,$
    at which the correction to the renormalized coupling is equal to the bare coupling, 
     we recover Eq.  (\ref{eq:FTK}), when approximating  $\tanh (x) \approx {\rm sign}(x)  ~{\rm  for} ~
     |x|>1$, and $ 0 $ otherwise, noting that
      the local density of states, the number of states per energy and volume,  can be written in terms of the eigen states of the conduction electrons as  
      \begin{equation}
      \rho({\bf r},\epsilon)  =  \sum_{n,\sigma} |\psi_n ({\bf r}) |^2 \delta (\epsilon-E_n).
      \end{equation} 
      In terms of the local density of states, we can thus  rewrite Eq. (\ref{eq:FTK}) as 
       \begin{equation}
\label{eq:FTKDOS}
1= \frac{V_a}{2}J \int_{0}^{D} d E \rho (E, {\bf r} )  \frac{1}{E-E_F} \tanh 
\left(  
\frac{E-E_F}{ 2 T_{\rm K} ({\bf r})} \right).
\end{equation}

 Since perturbation theory breaks down at temperatures of the order of $T_K$ nonperturbative treatment is needed to  
   derive lower temperature properties. 
 This is  possible with the 
 Wilson numerical renormalization group method\cite{Wilson75,krishna} and 
  analytically with the exact    Bethe-Ansatz method\cite{tsvelik,andrei}.  
Both methods show that the
  temperature $T$- and magnetic field $H$-dependence of the 
 free energy, and thus all thermodynamic observables, as well as transport properties
 like the resistivity scale with the Kondo temperature, depending only on the ratios 
 $T/T_K$ and $H/T_K.$
 Thus, thermodynamic observables like the magnetic susceptibility are for single Kondo impurities
 proportional to  known  universal  scaling functions   \index{scaling function}  of     $T/T_K$ and $H/T_K,$ 
  and it only remains to find the Kondo temperature for specific magnetic impurities in a metal. 
 The   low temperature phase  can thus be described by a 
 state where the magnetic impurity spins are screened by Kondo clouds formed by 
the conduction electrons whose effective   mass  is thereby enhanced, 
  but  still  forming  a Fermi liquid  \cite{Nozieres76}. 
  
  Therefore, let us first proceed to review the calculation of the Kondo temperature $T_K$. 
 For a clean metal  
 the   eigen states are plane waves with
  uniform intensity $|\psi_n ({\bf r})|^2 = 1/L^d,$  independent on position   ${\bf r}$.  
  For  smooth density of states $\rho(\epsilon_F) = \rho_0,$    we denote 
 the number of states per energy and spin as $N_0 =  V_a   \rho_0/2$, with $V_a = L^d/N$.  
 Then,  Eq. (\ref{eq:FTKDOS}) simplifies to 
    \begin{equation}
\label{eq:FTK0}
1= J \int_{0}^{D} d E N_0  \frac{1}{E-E_F} \tanh 
\left(  
\frac{E-E_F}{ 2 T_{\rm K}} \right).
\end{equation}
Noting that $\tanh (x) \rightarrow {\rm sign}(x)$ for $|x| \gg1$, 
and assuming that the Fermi energy is in the middle of the band $E_F = D/2$, 
we find $ 1 \approx J ~ 2 N_0 \ln ( D/T_K)$, 
yielding the Kondo temperature  \index{Kondo temperature} 
\begin{equation} \label{tk0}
T^0_{\rm K}  = c D \exp (- \frac{1}{2 N_0 J}),
\end{equation}
 where $c= 0.57$  is found by a more accurate integration of the $\tanh$-function. 
 Higher order corrections in $J$  lead only to pre-exponential corrections which  depend weakly  on $J$. Thus, the 1-loop result  $T^0_K$ yields already
  the dominant dependence on the exchange coupling $J$. 

         The Kondo effect can also  occur in semimetals, semiconductors and even  insulators, where  the density of states at the Fermi  
         energy is vanishing, when 
 the exchange coupling exceeds  a critical value $J_c.$ 
 At first sight, Eq.   (\ref{tk0}) seems to imply, that the Kondo temperature is vanishing when $\rho (\epsilon_F) =0$. However, 
  then the assumption of smooth density of states is   no longer valid and we need to
  start rather  from the general  self consistency equation, Eq. (\ref{eq:FTK}).
          In section \ref{gap} we will   therefore  consider and  review the  
         derivation of the Kondo temperature and $J_c$  for two generic cases: a)  when the Fermi level is  in  a pseudo gap and  b) when it is 
         in  a hard gap.

        In a real material there are  spatial variations of  local density of states 
         \index{local density of states} 
        $\rho({\bf r})$ and  exchange coupling $J$
        due to inhomogeneities and 
        disorder, both from nonmagnetic and magnetic impurities.
             According to  Eq.    (\ref{eq:FTK}) this results 
         in
          Kondo temperatures which  vary with  spatial position, $T_{\rm K } ({\bf r})$, since 
          the intensity $ |\psi_n ({\bf r})|^2 $ may vary spatially.
          Moreover the intensity of each state $ |n \rangle$ 
          at different  energy $E_n$ at the site of a magnetic moment may be different, making it a complex problem to evaluate the sum over all 
           eigen states. 
          In fact,      already in a weakly disordered metal   one finds that the 
     Kondo temperature is  distributed  with a   finite width\cite{jetpletters,micklitz06,prb2007}.     
      In section \ref{disorder} we will therefore consider and review   the Kondo effect in disordered systems in more detail.

\section{RKKY coupling between magnetic moments}

  A magnetic impurity never comes alone.  Thus, we need to consider what happens when more than one magnetic impurity
  is in the  metal. Naturally, the Anderson impurity  model Eq. (\ref{Anderson}) 
  can be extended to any number  $M$ of
   localized level sites,  summing over their $M$ positions $ {\bf r}_j$, 
\begin{eqnarray}
\label{AndersonM}
H &=& \sum_{n,\sigma} E_{n \sigma} \hat{n}_{n \sigma} +
 \sum_{j,\sigma}  \varepsilon_{d_j \sigma } \hat{n}_{d_j \sigma} + \sum_j U_j \hat{n}_{d_j +} \hat{n}_{d_j -}
 \nonumber \\ &+& \sum_{n, j, \sigma} \left( t_{n d_j} c^+_{n \sigma} d_{j 
 \sigma } + t_{d_j  n} d^+_{j  \sigma} c_{n \sigma } \right),
\end{eqnarray}
where the energy of localized levels $\varepsilon_{d_j \sigma}$,   onsite interaction $U_j$, and 
 hopping elements $ t_{d_j  n}$ may depend on the  positions 
  ${\bf r}_j$ with  $j= 1,...,M$. 
  Nothing prevents us, to perform again a  Schrieffer-Wolff transformation 
to  the  Kondo Hamiltonian \index{Kondo Hamiltonian} in the basis of the  singly occupied states of the $M$ magnetic moments, 
 which  yields 
\begin{equation} \label{KH}
H_K  = \sum_{n,\sigma} E_n \hat{n}_{n \sigma} +   \sum_{j, n,n'} J_{j, n n'}  \left[ S_j^+ c^+_{n +}  c_{n' -}  +  S_j^- c^+_{n -}  c_{n' +}  +  S_{j z}
( c^+_{n +}  c_{n' +} - c^+_{n -}  c_{n' -})
\right],
\end{equation}
      where   ${\bf S_j}$ is now  the spin vector  operator of the localized moment at position
  ${\bf r}_j$. 
Accordingly,   the matrix elements of the exchange coupling \index{exchange coupling}  depend on the positions  ${\bf r}_j$ as 
\begin{equation}
J_{j; n n'} =  t_{n d_j}\, t_{d_j n'} \left( \frac{1}{U_j+  \varepsilon_{d_j} - E_{n'} } +  \frac{1}{- \varepsilon_{d_j} + E_{n} }   \right).
\end{equation}
For the  symmetric Kondo model, we then get, 
\begin{equation} \label{HH}
H_K = \sum_{n,\sigma} E_n \hat{n}_{n \sigma} +  \sum_j J_j   \vec{{\bf S} }_j \vec{ {\bf s}} ({\bf r_j})
= H_0 +H_J,
\end{equation}
 with $J_j = 4 t_j^2/U_j >0.$
To derive the RKKY-coupling at finite temperature $T$,  let us consider   the thermodynamic potential $\Omega$
for  the Kondo model  Eq. (\ref{KH}). The correction $\Delta\Omega$
  due to the exchange interaction between  magnetic moments and the Fermi sea is 
  given by 
\begin{eqnarray}
\Delta\Omega=-T\ln\left\langle S \right\rangle = -T\ln \left({\rm Tr}  [S\cdot e^{- H_0/T}]/Z_0\right),
\end{eqnarray}
where $Z_0$ is the grand canonical  partition function of the Fermi sea, 
and $S$  the   correction factor due to the exchange interaction
term in the Hamiltonian, $H_J$,
$
S=\exp\left\{-\int_0^{1/T} H_{J}(\tau)d\tau\right\}.
$
Performing perturbation theory in   $J$ to second order we obtain 
\begin{eqnarray}
\label{omega}
&&\Delta\Omega=-\frac{1}{2} T\sum_{i,j;\alpha\beta\gamma\delta}
J_i J_j \int_0^{1/T}\int_0^{1/T}d\tau_1d\tau_2
\left\langle \vec{{\bf S}}_i{ \vec{ \sigma}}_{\alpha\beta}\vec{{\bf S}}_j{\vec{\sigma}}_{\gamma\delta} T_{\tau} [ c_{i\alpha}^{\dagger}(\tau_1)c_{i\beta}(\tau_1)
c_{j\gamma}^{\dagger}(\tau_2)c_{j\delta}(\tau_2) ] \right\rangle.\nonumber.
\end{eqnarray}
where $\langle ... \rangle = Tr [... \exp (- H_0/T)]/Z_0$. 
Here, we  assumed that  the conduction electron spins are not polarized, 
$\langle \vec{ {\bf s}} ({\bf r})\rangle=0.$
 Terms proportional to $\vec{{\bf S}}_i^2$ and $\vec{{\bf S}}_j^2$ yield only 
 corrections to the local energy, not to the  nonlocal interaction  $J_{\rm RKKY}$.
With  
Wick's theorem we can present the correlator in Eq. (\ref{omega}) in the form
$
-{\cal G}_{\beta\gamma}(i,j;\tau_1-\tau_2){\cal G}_{\delta\alpha}(j,i;\tau_2-\tau_1),
$
where
\begin{eqnarray}
{\cal G}_{\beta\gamma}(i,j,\tau_1-\tau_2)=-\left\langle T_{\tau}\left\{ c_{i\beta}(\tau_1)
c_{j\gamma}^{\dagger}(\tau_2)\right\} \right\rangle
\end{eqnarray}
is the Matsubara Green's function \cite{abrikosov}. 
 Since  we perform perturbation theory in $J$ to  2nd order,  only,
 and as long as there  are  no other spin dependent couplings in the Hamiltonian, 
  the propagator  
    ${\cal G}_{\beta\gamma}$   is proportional to
$\delta_{\beta\gamma} $ 
which allows us to  perform the summation over  spin indices in Eq. (\ref{omega})  explicitly to get 
$
\sum_{\alpha\beta}{\bf S}_i  \vec{ \sigma}_{\alpha\beta}{\bf S}_j  \vec{ \sigma}_{\beta\alpha}={\bf S}_i{\bf\cdot  S}_j,
$
 Thus, we find  in second order perturbation theory in $J$ that there is an 
  indirect exchange coupling  term in the Hamiltonian, the RKKY coupling 
   between the magnetic impurity spin operators $ \vec{{\bf S}}_i, \vec{{\bf  S}}_j,$ given by
\begin{eqnarray}
H_{RKKY}= \sum_{i,j} J_i J_j \chi_{ij}  \vec{{\bf S}}_i \vec{{\bf  S}}_j,
\end{eqnarray}
with the non-local, temperature dependent susceptibility matrix 
\begin{eqnarray}
\label{abr}
\chi_{ij}=-\frac{1}{2}\int_{0}^{1/T}{\cal G}(i,j;\tau){\cal G}(j,i;-\tau)d\tau.
\end{eqnarray}

\begin{figure}[t!]
 \centering
 \includegraphics[width=0.4\textwidth]{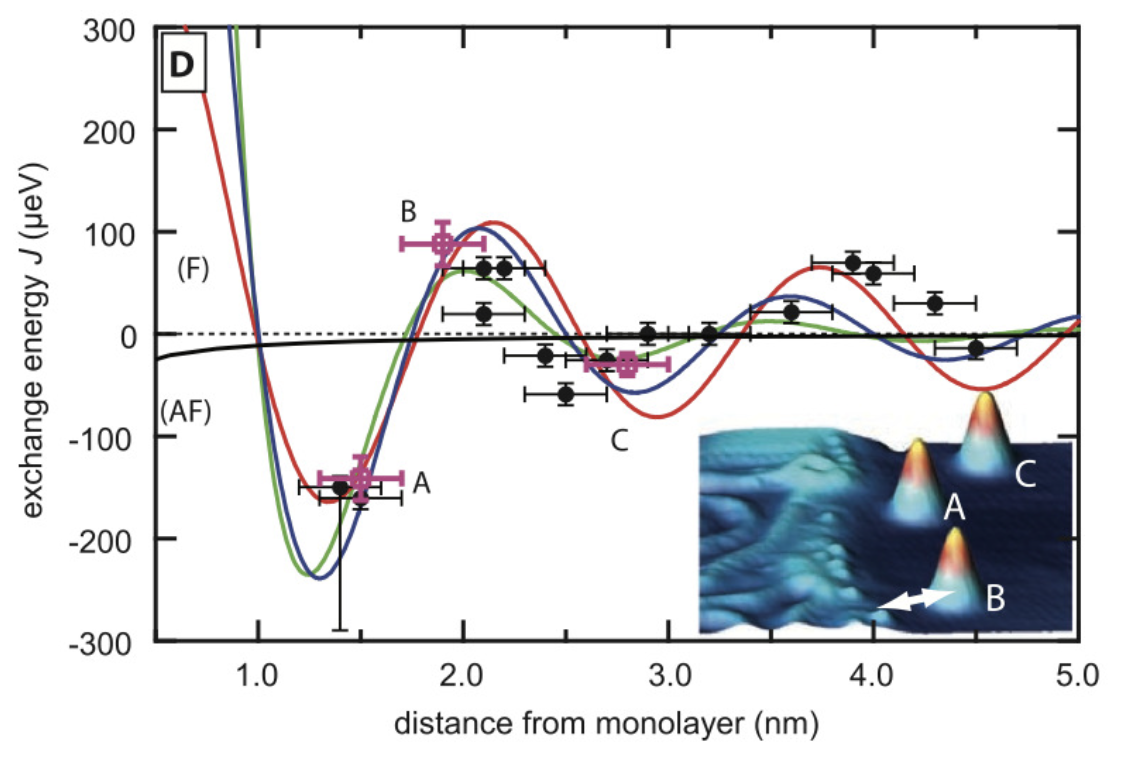}
 \caption{ 
Magnetic exchange interaction between adatoms on a monolayer stripe. Dots show measured exchange energy as  a function of distance from the monolayer, indicated  in the inset. Lines are fits to  the 1D, 2D and 3D RKKY-coupling \index{RKKY coupling} 
Eq. (\ref{JRKKY}). Figure taken from Ref. \cite{meier}.  
}
 \label{rkkystm}
\end{figure}

Writing the  Green's function
 in   the representation of eigen vectors $|n\rangle$,
\begin{eqnarray}
\label{abr2}
{\cal G}(i,j;\tau)=\sum_n \psi_n^*({\bf r}_i) \psi_n({\bf r}_j)e^{-(E_n-\mu)\tau}
\times\left\{\begin{array}{ll}-\left(1-f(E_n)\right), &\tau>0\\
f(E_n), & \tau<0 \end{array} \right.,
\end{eqnarray}
we find   the RKKY coupling  \index{RKKY coupling}  between the magnetic impurity spin operators $ \vec{{\bf S}}_i, \vec{{\bf  S}}_j,$
\begin{eqnarray}
\label{eq:Jij}
 J_{{\rm RKKY}}({\bf r}_{ij})  = J_i J_j \chi_{ij} = J_i J_j  \frac{  V_{a}^2}{4 \pi } 
{\rm Im} \int d E f(E) \sum_{n,l}
\frac{\psi^*_{n}({\bf r}_{i}) \psi_{n}({\bf r}_{j}) }{E - E_{n} +i \epsilon }
\frac{\psi_{l}({\bf r}_{i}) \psi^*_{l}({\bf r}_{j}) }{E - E_{l} +i \epsilon
},
\end{eqnarray}
where $V_{a}=L^d/N$.
Note that often when discussing the RKKY coupling, 
the magnetic impurity spins are treated classical. Then, the coupling has to be multiplied
 by $S(S+1)/S^2$ to account for  quantum fluctuations of the  magnetic impurity spins. 
   Here, we keep the quantum spin operators, since we  want to consider the competition with the Kondo effect, for which  quantum spin fluctuations  are essential. 
 We see that the RKKY coupling 
  depends not only on the local  intensities of the conduction electrons $ |\psi_{n}({\bf r}_{i})|^2$, but also on the phase difference between   the eigen functions at the different locations ${\bf r}_{i},{\bf r}_{j}$.
Inserting  plane-wave
states
$\psi_{n}({\bf r}_{i}) \sim \exp (i {\bf  k r}_i)$  
 into Eq. (\ref{eq:Jij})  one finds at large distances    $k_{\rm F} r_{ij} \gg 1$  the 
 RKKY coupling   \index{RKKY coupling} \cite{aristov},
 \begin{equation} \label{JRKKY}
 J^{0}_{{\rm RKKY}}({\bf r}_{kl})
\rightarrow   -  c_d N_0 J_{i} J_{j} \sin (2 k_{\rm F} r_{ij} + d \pi/2)  \frac{V_a }{ r_{ij}^{d}},
\end{equation}
 where $ r_{ij}=| {\bf r}_{i}-{\bf r}_{j} |$, with  dimension $d$,  Fermi wave numbe $k_{\rm F},$ $V_a=V/N$ and $m$ the effective electron mass.  Here $c_2 = 1/\pi, c_3 = 1/(2 \pi)$, and $N_0 = 1/D$.
 Here,  $N_0  = V_a \rho_0/2$ is the number of states per energy and spin with 
  total density of states  (including the factor 2 for spin)  $\rho_0 = m/\pi$ in $d=2$ dimension and  $\rho_0 = m k_{\rm F}/\pi^2$
  in $d=3.$ 
In Fig.  \ref{rkkystm}   results for the coupling between two magnetic adatoms on a metal surface are plotted for various distances ${\bf r},$
as extracted from  spin dependent scanning tunnelling microscopy measurements\cite{meier}.

\section{ Spin competition: the Doniach diagram  }
\label{dd}

\begin{figure}[t!]
 \centering
 \includegraphics[width=0.4\textwidth]{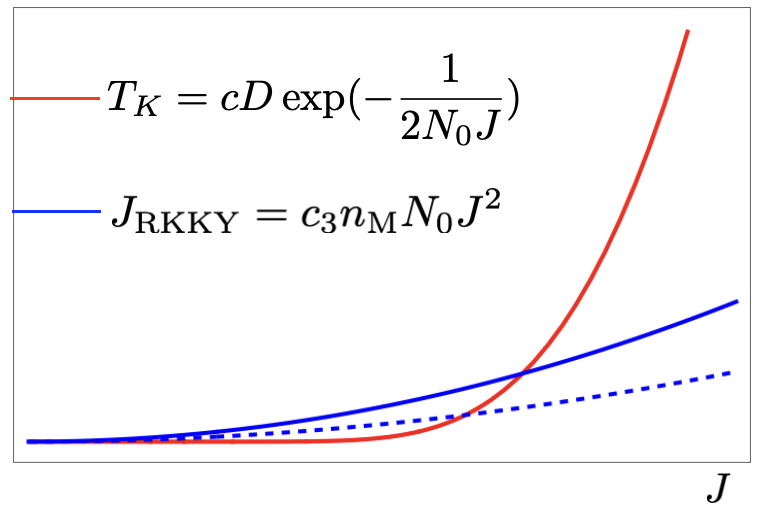}
  \includegraphics[width=0.42\textwidth]{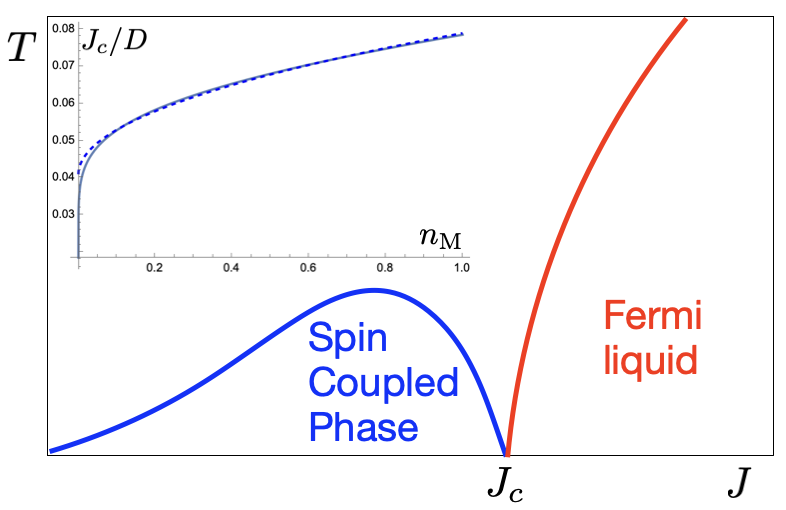}
 \caption{ Left:  Kondo temperature $T_{\rm K}$ with $c= 0.57$ (red) and 
  RKKY coupling  $J_{\rm RKKY}$  with $c_3 =1/(2 \pi)$ for (dense, dilute) magnetic impurities, $n_M=(1,0.5) \equiv$ (blue,blue dashed)  
  as function of exchange coupling $J$.  Right:  Doniach diagram\index{Doniach diagram}, 
 qualitative sketch of  transition temperature to  a spin coupled state (blue), and to the 
  low temperature Fermi liquid (red), as function of $J$.  The critical coupling 
  $J_c/D$, 
  Eq. (\ref{jc}), is plotted in the inset as function of magnetic moment density $n_{\rm M}$ (blue line), together
   with the fit 
  $J_c/D \approx 0.041 + 0.038 \sqrt{n_{\rm m}}$(dashed blue line).  }
 \label{dpd}
\end{figure}

Knowing 
 the Kondo  temperature $T_{\rm K}$   and the  
 RKKY coupling 
 we can now study  their competition as function of the local exchange coupling $J$ and concentration 
  of magnetic moments $n_{\rm m}$.
 The amplitude of the oscillatory RKKY coupling 
  Eq. (\ref{JRKKY}) can be rewritten as  $J^0_{RKKY}/D = c_d  J^2 n_{\rm m}$.
   Noting  that the coupling is dominated by nearest neighboured magnetic moments, we
  write   it in terms of the density of magnetic moments
    $n_{\rm M} = V_a/R^d,$  where $R$ is the average distance between next neighboured magnetic moments.  
   In  Fig. \ref{dpd}   (left) we plot  both energy scales in  $d=3$ dimensions.
 For the RKKY-coupling we plot it  both   for 
   a dense system of magnetic impurities 
    $n_{\rm M} = 1$ (blue), and for a more dilute case,  $n_{\rm M} = 0.5$ (dashed blue). 
  
Thus, we see that there is a critical coupling $J_c$ below which the RKKY coupling exceeds the 
 energy scale for Kondo screening $T_{\rm K}$, so  that the magnetic impurity spins can be coupled 
 with each other. That critical coupling  $J_c$ is seen to increase with   the concentration  of  magnetic moments.
  Solving the nonlinear equation analytically, we find 
  \begin{equation} \label{jc}
  J_c = - \frac{D}{ 4 W(-1,- \frac{1}{4}\sqrt{\frac{c_d}{c} n_{\rm m}})},
  \end{equation}
  where $W(k,z)$ is the $k$-th branch of the Lambert W-function, also known as ProductLog-function,  plotted for $d=3$ in the inset of  Fig. \ref{dpd}   (right), blue line. 
   We find  for the whole range of concentrations $0<  n_{\rm m} <1$, 
   $J_c/D \approx 0.041 + 0.038 \sqrt{n_{\rm m}}$ a good fit (dashed blue line). 

   Doniach argued in Ref. \cite{Doniach77}
 that  the critical 
    coupling  $J_c$ marks a quantum phase transition between 
    a   heavy fermion
  \index{heavy fermions} state   and an  ordered, typically antiferromagnetic, phase  \index{ordered phase} 
 in the Kondo lattice limit $n_{\rm M} =1$, where nearest neighbour RKKY coupling is antiferromagnetic.
This  gives a good description of  quantum phase transitions   \index{quantum phase transition}   in  
heavy fermion materials
 that contain rare earth elements  like Ce, Sm, and Yb or actinides  like  U and Np,
where local magnetic moments originate from localized f-orbitals and 
antiferromagnetic order is observed  at sufficiently low temperature. As pressure or external magnetic field is changed, 
  the N\'eel temperature $T_N$ reaches a maximum, before 
 it is suppressed   at the quantum critical point. This has been 
 measured in detail for  Cerium compounds, such as CeAl$_2$, CeAg or CeRh$_2$Si$_2$,
 as well as 
in  YbRh$_2$Si$_2$ 
 under pressure
 and in a magnetic field\cite{Seiro},
  for a review see\cite{Coleman2007}.
           \begin{figure}[t!]
 \centering
 \includegraphics[width=0.5\textwidth]{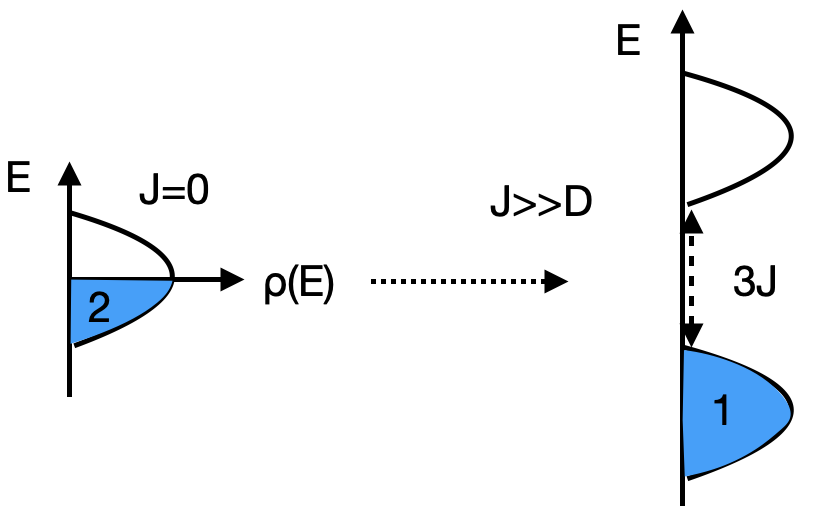}
 \caption{Left: Sketch of typical  density of states $\rho(E) $ of a metal as function of energy E.
 The band of width $D$ is half filled   with $N$  electrons (blue).
 Each state is doubly spin degenerate as indicated by $2$.
  Right: Density of states in the presence of  a lattice of  $N$ magnetic moments, 
  coupled by   strong  antiferromagnetic exchange coupling $J>D$ to the conduction electron spins, 
 forming $N$ non degenerate Kondo singlet states, as indicated by  1, thereby enlarging the Fermi surface. 
  There is a charge transfer gap  $3J,$  making that system for  $J>D$ a Kondo insulator  \index{Kondo insulator} .  
  }
 \label{ki}
\end{figure}

 In order to derive that quantum phase transition  at $T=0K$ 
  one needs to  find the ground state of  the  Anderson model with a finite number of $M$  Anderson impurity sites, Eq. (\ref{AndersonM})
  or, alternatively, solve the Kondo model of $M$ impurity spins,  Eq.  (\ref{HH}) as coupled to the conduction band by the 
  antiferromagnetic 
  exchange couplings $J_j = 4 t_j^2/U_j>0$
   at the $M$ impurity sites $j=1,...,M$. 
   In a dense Kondo lattice, where $n_M=1$ and  the number of impurity spins 
    $M$ equals the number of occupied conduction  band states $N$.  
    
    Let us   start by looking at a simpler state, a Kondo insulator  \index{Kondo insulator} , which can for example  form  when the  uniform   coupling is strong $J= J_j\gg D$, exceeding  the conduction band width $D$.
Then,  the exchange coupling $J$  is so strong that  each impurity spin localizes one of the conduction electrons, so that the 
      ground states is given simply by a product of singlet states,  $|\psi_0 \rangle = \sum_{j=1}^N  | 0_i \rangle$, where 
      $ | 0_i \rangle  = 1/\sqrt{2}( | \uparrow_{d i}  \rangle   | \downarrow_{c i}  \rangle  -  | \downarrow_{d i}  \rangle   | \uparrow_{c i}  \rangle  ) $,  is the singlet state formed by the impurity spin  (indexed by $d$)  and a conduction electron spin  (indexed by $c$) at site $i$.   The  ground state energy is then given by $E_0 = - N  (3/2) J$. 
        The lowest   spin excitation energy gap $\Delta E_s$ is  obtained by exciting one of the spin pairs  to a triplet state of energy $J/2$, thus $\Delta E_s = 2 J.$
        However, there is also a finite charge transfer gap  $\Delta E_q,$ which is obtained by transferring one of the conduction electrons 
         from site $i$ to site $j$. Consequently, at site $i$  the impurity spin is left alone  $ | \uparrow_{d i}  \rangle $, breaking up the singlet state $ | 0_i \rangle $, 
          and shifting its energy to $E_i =0$, 
          while at site $j$ the singlet state is also broken up to  accommodate the second electron, exciting the state to 
          $( | \uparrow_{d i}  \rangle   | \downarrow_{c i}  \rangle   | \uparrow_{c i}  \rangle $, with energy  $E_j =0$, since the spins of the two conduction electrons compensate each other to $s_{\rm tot ~j} =0$, so that the magnetic impurity spin at site $j$  cannot couple to them. As a consequence, a transfer of a single electron 
           from site $i$ to site $j$ costs  in total $ \Delta E_q = 2 (3/2) J$. Thus,  the exchange coupling $J$ to localized 
            magnetic moments prevents charge transfer, opening a large  gap 
            $ \Delta E_q = 3 J$. Taking into account the finite band width $D$ due to the dispersion of the conduction electrons, each of the $N$
              Kondo singlets  is formed rather  by electrons in   superpositions of conduction band states. Thus, these Kondo clouds overlap strongly in space. 
                To accommodate all $N$ conduction electrons in these $N$ Kondo singlets, the Fermi surface  expands to embrace  all states in the conduction band, 
               as  compared to the half filled conduction  band   with  doubly occupied states, which is the ground state without the exchange coupling $J$, see 
               Fig. \ref{ki} and 
               the discussion 
                in  Refs.  \cite{Nozieres1998}, \cite{Coleman2007}. 
            Above this ground state of Kondo singlets, the gap 
            $\Delta E_q = 3J$ opens,   the energy  needed to transfer one electron from one   Kondo singlet   to another, 
               making the system an insulator. 
               In fact, this is the mechanism for the formation of a Kondo insulator, which has been  experimentally observed, first   in $ SmB_6$\cite{KI}.

\begin{figure}[t!]
 \centering
 \includegraphics[width=0.5\textwidth]{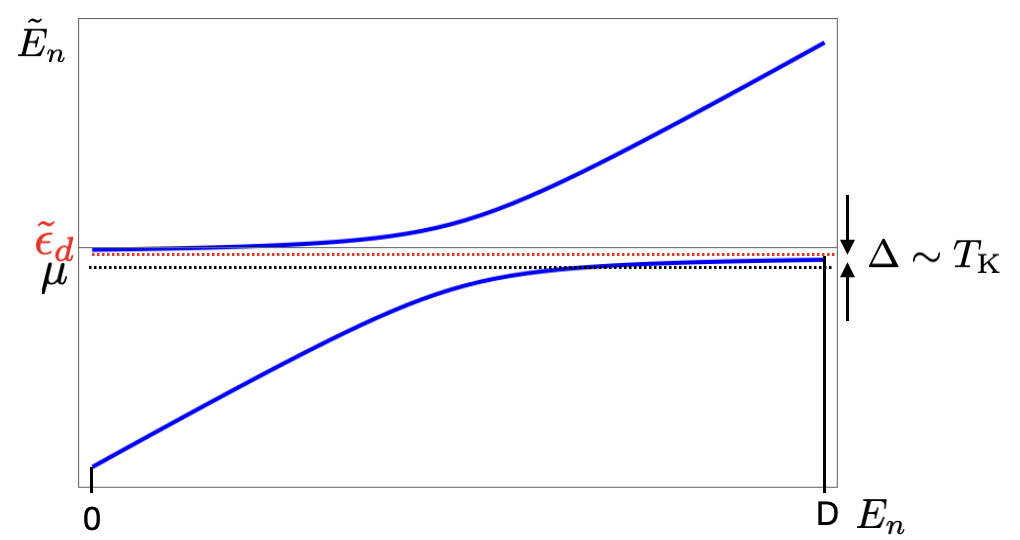}
 \caption{ Quasi particle eigen energies $\tilde{E}_n $ as function of bare  eigen energies  without Kondo coupling   $E_n.$  $\mu$ is the chemical potential,   the energy level of the localized level 
   becomes shifted upward into the gap, $\tilde{\epsilon}_d.$  The indirect  gap $\Delta$ is
    of  the order of $T_{\rm K}$,  
  }
 \label{hf}
\end{figure}

              When the exchange coupling is smaller than the band width $J<D$, the ground state is no longer a simple  product of Kondo singlets.
               One way to derive the ground states then is 
     by a mean field treatment of a generalised Kondo lattice  Hamiltonian 
 with degeneracy $N_K \gg 1$, the Coqblin-Schrieffer Hamiltonian
  \cite{Coqblin69},  performing  a $1/N_K$-expansion, as done  first in Refs. 
\cite{Read83}  and \cite{Auerbach86}.  Thereby one finds the  quasi particle 
  \index{quasi particle} 
eigen energies $\tilde{E}_n $ as function of the eigen energies of the 
 conduction band without Kondo coupling,   $E_n,$  
\begin{equation}
\tilde{E}_n = \frac{1}{2} (E_n + \tilde{\epsilon}_d) \pm  \frac{1}{2} \sqrt{ (E_n - \tilde{\epsilon}_d)^2 + 4 V^2}.
\end{equation}
as plotted in Fig \ref{hf} as function of $E_n$
Thus, there opens a gap  $\Delta = T_{\rm K}$,   relating  the mean field order parameter $V$ to the Kondo temperature  $T_{\rm K}$  by 
 $4 V^2  = D  T_{\rm K}$, when $\tilde{\epsilon}_d \approx D/2.$ 
 The density of states above and below the gap 
 is seen to be strongly enhanced. The chemical potential is located in the lower band, so that the exchange coupling transformed the  Fermi sea of conduction electrons 
   to a Ferm sea of heavy holes. The energy level of the localized level $\epsilon_d$ 
   becomes shifted upward into the gap, $\tilde{\epsilon}_d.$ 
  With the invention of dynamical mean field theory, exploiting the fact that mean field theory becomes exact in infinite dimension
  $d \rightarrow \infty$
\cite{Metzner}, another route to solve the Kondo lattice model  and  the periodic  Anderson model 
opened, which allows the calculation of self energies 
\cite{Jarrell}, 
\cite{Georges} and of the resistivity. Thereby, the decay of the  resistivity  was shown as coherent heavy fermions 
form at low temperature
\cite{Schweitzer}, in agreement with  experiments on heavy fermion compounds, as  reviewed above. 

However, in order to allow the study of  the competition between  the Kondo screening and the RKKY-coupling 
both approaches need to be modified. 
By adding the RKKY-coupling between the magnetic moments to the Kondo lattice hamiltonian, 
 the quantum phase diagram can be studied in mean field theory, when  combined with the $1/N_K$-expansion
\cite{Theumann2001,Coqblin2008,Magalhaes2010,Magalhaes2012}.

A system of two magnetic impurities in a metal  has  been studied in detail   with nonperturbative methods like the numerical renormalization group
 \cite{Jones,Bulla2008,Eickhoff18,MitchellBulla,Eickhoff20}. 
 Such a system has also been realised  experimentally by Co atoms on a gold surface  
 and studied varying their distance 
  with scanning tunnelling microscopy\cite{Bork2011}.
One finds a crossover between a state where both magnetic moments are   Kondo 
screened by the conduction electrons and a state where the  impurity spins are coupled. Depending on their distance, 
they  form  either a singlet, enforced  when RKKY coupling is antiferromagnetic, or a triplet state, when the coupling is ferromagnetic. 
    Building on these studies, DMFT has  been extended to Cluster-DMFT, where the exact results for a cluster of few spins, 
     in particular two magnetic impurities are used to
     enhance the DMFT and to 
      derive the phase diagram of the Kondo lattice model with RKKY coupling 
    \cite{Delft23}. 
    
  Further insights into this competition comes from exact analytical  results  for the  1D  Kondo lattice and the 1D periodic Anderson model
  employing the     bosonization   \index{Bosonization} technique, see Refs. 
    \cite{Fujimoto97} and \cite{Tsunetsugu97}  and references therein. At half filling a Kondo insulator is found, and both the spin and charge transfer gaps have been derived \cite{Fujimoto97}.

Recently, Nejati et al. extended the 
renormalization group equations for  a  Kondo  lattice  incorporating self consistently the RKKY coupling between magnetic moments \cite{Nejati2017}   \index{self consistent renormalization group} . 
Thereby they could show that the Kondo temperature is decreased as the exchange  coupling $J$ is decreased,
as   found with diagrammatic methods in Refs.  
\cite{Matho72,Tsay73,Tsay75}. Furthermore, it was found in Ref. \cite{Nejati2017} that
 the Kondo screening is quenched at  a critical
 coupling $J_c$. Since this approach to the spin competition problem is very   insightful  let us review it    in the remainder of this section.



 The renormalization of the effective coupling $\tilde{J}(\Lambda)$
 at energy $\Lambda$, above and below the Fermi energy, Eq. (\ref{rg}), is modified by the RKKY coupling as derived first  in Ref. \cite{Nejati2017} and generalised  in Ref.   \cite{Park2021}  to account for an energy dependent local 
 density of states $ \rho(E,\bm{r}_i),$ yielding
 \begin{eqnarray} \label{rgnm1} 
&& \frac{d \tilde{J}_i}{d \ln \Lambda} = -  \tilde{J}_i^{2} \frac{V_a}{2} ~\sum_{\alpha=\pm} 
\rho(\mu+ \alpha \Lambda,\bm{r}_i)  
 \nonumber \\
&&+    \frac{4  }{\pi} \tilde{J}_i^{2} J_i^0  ~\sum_{\alpha=\pm} 
  \sum_{j \neq i} J_j^0 {\rm Im} [e^{i {\bf k}_F {\bf r}_{ij} } \chi_c ({\bf r}_{ij},\mu+ \alpha \Lambda) 
   G^R_c ({\bf r}_{ij},\mu+ \alpha \Lambda) \chi_f ( {\bf r}_{j},\mu+ \alpha \Lambda ) ],
   \end{eqnarray}
  where  $\Lambda$ is the effective band cutoff  for the renormalization group flow. 
 While the first term on the right hand side is the well known  
  1-loop RG for the Kondo problem with energy  dependent  density of states, Eq. (\ref{rg})
\cite{Suhl65,Zarand96},  the second term describes the  correction due to RKKY-coupling. 
Here, $\chi_f ( {\bf r}_{j},E) $ is
the spin susceptibility of the magnetic moment  
 at  site ${\bf r}_{j}$.
$ G^R_c ({\bf r}_{ij},E)$ is the retarded   conduction electron propagator from  site 
 ${\bf r}_{i}$ to ${\bf r}_{j}$  and we defined the distance vector 
${\bf r}_{ij} = {\bf r}_{i} - {\bf r}_{j}$.
$\chi_c ({\bf r}_{ij},E) $ 
denotes  the conduction electron correlation function between sites ${\bf r}_{i}$ and ${\bf r}_{j}$.
Solving Eq. (\ref{rgnm1})  we can thus derive   the position dependent Kondo temperatures  for  a given configuration of 
magnetic moments.

When the magnetic moment density  $n_M$ is not too large,
 $\chi_f ( {\bf r}_{j},E) $  can be approximated  by the
 Bethe-Ansatz solution   for a single Kondo impurity\cite{tsvelik,andrei}.
  In
   Ref. \cite{Nejati2017} this approximation has been used.  Then, 
only   its  real part contributes, as given by
  ${\rm Re} \chi_f ( {\bf r}_{j},\mu+D)  =  \mathcal{W}/(\pi T_{Kj} \sqrt{1+D^2/T_{Kj}^2}).$ Here, $\mathcal{W}$ is the Wilson ratio  \index{Wilson ratio}. 
  $T_{Kj}$ is the Kondo temperature  at site ${\bf r}_{j}$.
Since it is well known that
       the energy dependence of the density of states
        changes the Kondo renormalisation \cite{pseudogap}, 
         it is in general 
          important to keep the energy dependence of all functions and not to replace it with  their value at the chemical potential, 
         when the density of states is strongly varying with energy, as in the presence of a pseudo gap, or in disordered systems. 
         
      But, let us  first consider  the simpler case of 
 magnetic moments in a clean  metal, with slowly varying density of states.  
Then,  we can furthermore assume that
   all conduction electron properties, the local density of states,
   the propagator $ G^R_c ({\bf r}_{ij},E)$ and  the
   correlation function $\chi_c ({\bf r}_{ij},E) $ depend only
    weakly on energy, and therefore can be replaced by its value
     at the chemical potential $\mu$, as has been done in  Ref. \cite{Nejati2017}. 
Then, we can define the effective Kondo coupling $g_i = N(\mu) J_i$  of the Kondo impurity at site ${\bf r}_i,$ where $ N(\mu) = V_a \rho(\mu)/2$,
and find the  renormalization group equation for $g_i$, as modified by the RKKY coupling\cite{Nejati2017},
\begin{equation}  \frac{d g_i}{d \ln \Lambda} = - 2 g_i^{2} \Big( 1 - y_i g_{0}^{2} \frac{D}{2T_{K}} \frac{1}{\sqrt{1 + (\Lambda/T_{K})^{2}}} \Big) , \end{equation}
where  $D$ is the bare bandwidth and $g_0 = N(\mu) J^0$ is the bare, unrenormalized Kondo coupling. $y_i$ is the effective dimensionless RKKY interaction strength at site ${\bf r}_i$, given by \cite{Nejati2017}
\begin{equation} y_i = - \frac{8 W}{\pi^2 \rho(\mu)^2} {\rm Im} \sum_{j \neq i} e^{i {\bf k}_F {\bf r}_{i j}} G^R_c ({\bf r}_{i j}, \mu) \Pi ({\bf r}_{i j}, \mu) ,
\label{y}
\end{equation}
%
%
where $W$ is the Wilson ratio as determined by the Bethe Ansatz solution of the Kondo problem\cite{tsvelik,andrei}. $G^R_c ({\bf r}_{i j})$ is the single particle  propagator in the conduction band from site ${\bf r}_{i}$ to ${\bf r}_{j}$.
 The summation is over all other magnetic moments at positions ${\bf r}_{j}$.
 $\Pi ({\bf r}_{i j}, \mu)$ is the RKKY  correlation function of  conduction electrons between sites  ${\bf r}_{i}$ and ${\bf r}_{j}$.
  $y_i$ is found to be always positive \cite{Nejati2017}, while the RKKY correlation function can be positive or negative.

It is interesting to observe that the effective Kondo interaction renormalized by the RKKY interaction is a function of $\Lambda/T_{K}$, where $\Lambda$ is the renormalization group energy scale and $T_{K}$ is the renormalized Kondo temperature to be determined self-consistently.

\begin{figure}[t!]
 \centering
  \includegraphics[width=0.42\textwidth]{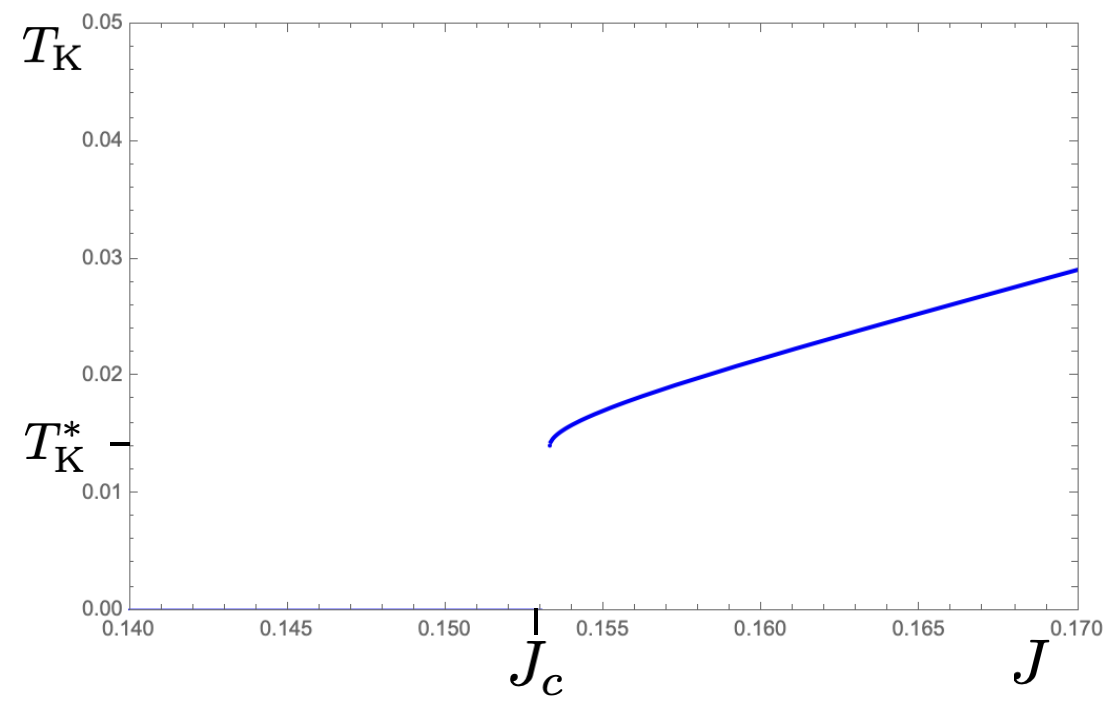}
 \caption{   Kondo temperature as function of $J$  in units of $D$,  $T_{\rm K}(J)$, Eq. (\ref{tky}), 
 as modified by RKKY coupling for  $y= 0.6/k <y_c$. 
 Note that $T_{\rm K}(J)$  jumps discontinuously from $T_K^*$ to zero  at 
  $J_c$.  }
 \label{dpdm}
\end{figure}

For two magnetic moments in a clean system, where the
bare couplings $g_0$ are the same at both sites, and $y_i=y$,
one can solve this differential equation to obtain \cite{Nejati2017}
\begin{equation} \frac{1}{g} - \frac{1}{g_{0}} = 2 \ln \Big( \frac{2 \Lambda}{D} \Big) - y g_{0}^{2} \frac{D}{2T_{K}} \ln \Big( \frac{\sqrt{1 + 
(\Lambda/T_{K})^{2}} - 1}{\sqrt{1 + (\Lambda/T_{K})^{2}} + 1} \Big) . \end{equation}
When the energy scale  $\Lambda$ coincides with the Kondo temperature, i.e., $\Lambda \rightarrow T_{K}$, the correction to  the effective Kondo interaction  is large. Therefore, setting  $g(T_{\rm K}) = \infty$
 we get the  self-consistent equation for the effective Kondo temperature as a function of the RKKY interaction,
\begin{equation}
\label{tky}
T_{K}(y,g_0) = T^0_{K}(g_0)  \exp (- y k g_{0}^{2} \frac{D}{T_{K}(y)} ),
 \end{equation}
where $T^0_{K}(g_0) = c D \exp (-1/(2 g_0))$ is the bare Kondo temperature in the absence of the RKKY interaction and the numerical constant is $k = \ln (\sqrt{2} + 1)$. 
Its solution is 
\begin{equation}
\label{tky}
  T_{K} (y,g_0) =- \frac{y k g_0^2 } {W (-y k g_0^2 /T^0_K(g_0) )},
 \end{equation}
for $y<y_c$ with the critical
 coupling   \index{critical coupling} \cite{Nejati2017}
\begin{equation} \label{yc}
y_c= T^0_{K}/(k~ e g_0^2 D ).
\end{equation}
Noting that the coupling $y$  is related to the magnetic moment density as  $ y \sim  n_{\rm M}$, and  that $g_0 = N_0 J$, 
we find that  the exchange coupling has to exceed 
  the critical coupling $J_c (n_{\rm M})$.  
Thus, it agrees with the   result obtained above, when 
  using  the Doniach argument, Eq. (\ref{jc}), up to a numerical constant of order 1. 
  As  the  exchange coupling $J$ is diminished  toward that critical value, $J_c$,  the Kondo temperature 
$T_K$ becomes diminished continuously, as plotted in Fig. \ref{dpdm}. At the critical value, however, 
 it is found to take a finite value $T_K^*=T_{K c}(J_c)  = e^{-1}T^0_{K}(J_c),$ about one third of its value without the RKKY coupling, before it jumps to zero at smaller $J$.

For two magnetic   Co atoms on a gold surface   such a suppression of $T_K$ was observed experimentally in Ref. 
  \cite{Bork2011}  
 at varying distance $R$, as measured in the width of the tunnelling peak 
  with scanning tunnelling microscopy\cite{Bork2011}.
In that case, one finds a crossover between a state where both magnetic moments are   Kondo 
screened by the conduction electrons and a state where the  impurity spins form a singlet, enforced  when RKKY coupling is antiferromagnetic. 
  
  Applying  that to  a system of dense magnetic moments, like  heavy fermion materials, 
   this result is remarkably different from     the  Doniach diagram, Fig.  \ref{dpd},  where it was
 assumed that both the Kondo temperature and the 
  critical temperature  $T_c$ on the spin coupled side of the transition would decay continuously towards the critical 
   point $J_c$. However, to conclude on the nature of the quantum phase diagram one would have to include self consistently the change in the spin polarisation function in the derivation due to an ordering transition 
    of the unscreened or partially screened magnetic moments or by  a spin wave  instability of the conduction electrons. 
  
The result Eq. (\ref{tky})  also implies  that by taking into account the RKKY-coupling, the Kondo temperature 
  becomes dependent  explicitly  on the distance  $R$ between the magnetic moments, and thereby on the density of magnetic moments 
$n_{\rm M} $.


  However, when  the magnetic impurity concentration   $n_{\rm M} $ is lowered, not only is $J_c \sim   \sqrt{n_{\rm m}}$ diminished, and thereby the parameter range of the ordered phase reduced, 
    but the positions of magnetic moments   become distributed randomly. 
      Thus, for  $n_{\rm M} <1$, the distance between  magnetic moments $R$ is random. 
            Thereby, 
              both the  sign and amplitude  of the RKKY coupling 
             is randomly distributed.  This may give rise to 
         the appearance of a richer quantum phase diagram with a 
          spin coupled  phase without long range order,  such as a {\it spin glass} state\cite{SK},
       which  competes with an ordered state\cite{Theumann2001}. 	
       In metal wires with dilute magnetic impurities, such as Ag$_{1-x}$Mn$_x$, 
       a transition from a Kondo phase to a spin glass phase has been detected in transport experiments, as the Mn concentration 
       $x$ is enhanced\cite{Forestier}. 
  Spin glass phases have also been found in 
   alloys with rare earth elements, such as  CeNi$_{1-x}$Cu$_x$\cite{Gomez}, where 
    the competition between Kondo and RKKY coupling is studied 
     as function of $x$:
     CeCu  ($x=1$)  is 
     at low temperature an antiferromagnet and  the alloy remains one up to $x=0.7,$
     while CeNi ($x=0$) is a  heavy fermion material.  Thus,  lowering $x$ corresponds to 
      an increase of the local Kondo coupling $J$, inducing a Doniach like quantum phase transition. However, at intermediate values of $x$, 
       disorder is relevant, and 
     spin glass  behaviour  is found\cite{Gomez}, as reviewed and modelled 
     in 
     \cite{Magalhaes2012}.  Similar successions of  quantum phase transitions between 
      heavy fermion, spin glass and ordered phases have been found in 
      CeRh$_x$Pd$_{1-x}$  as function of $x$\cite{Westerkamp}. 
We will consider the effect of disorder on the competition between Kondo screening and 
 RKKY coupling  in section 7, where we find that new  effects introduced by randomness, like 
 Anderson localization and multifractality have to be taken into account in strongly disordered systems with magnetic moments, 
 which profoundly change  the quantum phase diagram. 
  
  In the next section we consider the effect of    strongly varying  density of states on both the Kondo screening and the RKKY-coupling and thereby on the quantum phase diagram.
  
\section{  Spin competition in  presence of a   spectral (pseudo) gap} 
\label{gap}

In semiconductors and insulators the density of states at the Fermi energy is vanishing. 
 At first sight, Eq.   (\ref{tk0}) seems to imply, that the Kondo temperature is vanishing in such a situation, when $\rho (\epsilon_F) =0$.
However, 
 the assumption of smooth density of states is  no longer valid and we need to
  start   from the general  self consistency equation, Eq. (\ref{eq:FTK}).
  We find that the Kondo effect occurs provided the exchange coupling $J$ exceeds a critical value $J_c$. Let us review the derivation of the Kondo temperature and $J_c$ for two generic cases when the Fermi level is in a pseudogap and when it is in a hard band gap.

\subsection{Band insulator, semiconductor} 
\label{gap}

        Here, we derive   the Kondo temperature in  a band insulator   \index{band insulator}  with a gap $\Delta$,  where
       the Fermi level  is in the middle of the gap, as sketched in Fig. \ref{tki} (left), by
         inserting the gapped density of states into Eq. (\ref{eq:FTKDOS}).
         For  a small gap
        $ \Delta < T_{\rm K} $,
       assuming that the density of states is constant  and the same in the upper and lower band, $\rho_0,$ 
        the functional dependence of the Kondo temperature on the exchange coupling $J$ remains  the same as in a metal, 
        $T_{\rm K} \approx c(\Delta) \exp (-1/(2 N_0J ))$, where $N_0$ is the number of states per energy and spin, 
        but the pre-factor $ c(\Delta) < c$ is  diminished compared to   a metal, $c = c(\Delta =0) \approx 0.57$. 
         When the gap is larger $ \Delta > T_{\rm K},$ the functional dependence on $J$ changes. 
         Integration of  Eq. (\ref{eq:FTKDOS}),  using that $\tanh (x) \approx 1- 2 \exp(-2 x)$ for $x>1$, 
         we find the Kondo temperature 
         as a solution of the equation
         \begin{equation}
         \frac{\Delta}{4}  \left( \ln \frac{D}{\Delta} - \frac{1}{2 N_0 J} \right)  =  T_{\rm K} \exp (- \frac{\Delta}{T_{\rm K}}).
         \end{equation}
         We see that, only when the left side is positive,  can there be a real solution for $ T_{\rm K} $. Thus,
          the Kondo temperature can only be finite when $J> J_c(\Delta)$, with 
          critical exchange coupling   \index{critical exchange coupling} 
          \begin{equation} \label{jcd}
          J_c^{\Delta} = \frac{1}{2 N_0}
 \frac{1}{\ln (D/\Delta)}.         \end{equation}
  As $J \rightarrow J_c^{\Delta} $ the Kondo temperature is found to decay
   continuously to zero as 
     \begin{equation} \label{tki}
       T_{\rm K}  = \frac{ \Delta }{2} \frac{1}{ W (4 N_0 J/(J/J_c^{\Delta}-1)   ) } .
           \end{equation}
 For $J \gtrsim J_c^{\Delta}$ that decay can be approximated   as 
    $
     T_{\rm K}  \approx \Delta/2 / \ln (4 N_0 J/(J/J_c^{\Delta}-1)  )$. 
Thus, for $J>J_c^{\Delta}$ magnetic moments are
Kondo screened for temperatures below $T_{\rm K}(J)$, in spite of
 the large band gap $\Delta > T_{\rm K}  $. 
 For $J<J_c^{\Delta}$ all magnetic moments remain unscreened. 

        \begin{figure}[t!]
 \centering
  \includegraphics[width=0.2\textwidth]{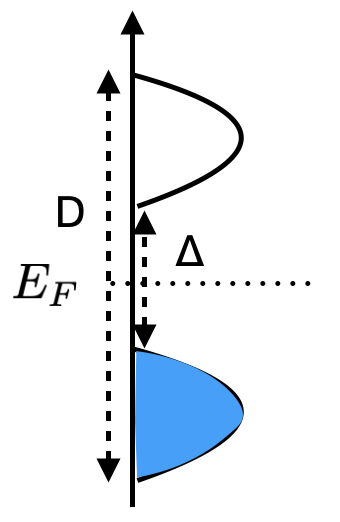}
    \includegraphics[width=0.5\textwidth]{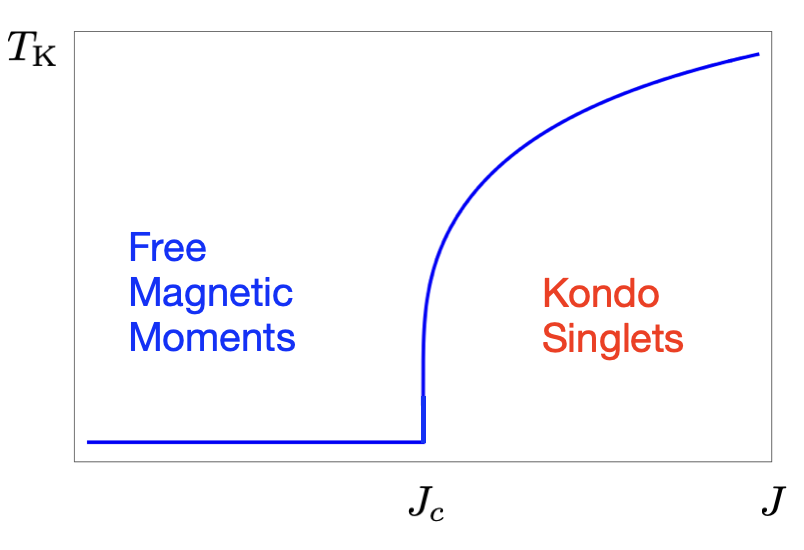}
 \caption{ Left: schematic density of states  with gap $\Delta$, total bandwidth $D$,  Fermi energy $E_F$ in the middle of the gap. 
 Right:   Kondo temperature $ T_{\rm K}(J)$ as function of $J$ in units of $D$,  Eq. (\ref{tki}).
 Note that it decays continuously to zero  at  the critical coupling 
  $J_c(\Delta)$. Eq.  (\ref{jcd}).  }
 \label{tki}
\end{figure}
Since  RKKY couplings, Eq. (\ref{eq:Jij}), are dominated by the density of states at the Fermi energy, 
in an insulator they are small and  decay exponentially with distance $R$. 
However, when an insulator or semiconductor is doped, 
  there exist other forms of magnetic coupling between magnetic dopants.
   For spin-1/2- dopants, like P in Si, 
   they are known to be coupled by  antiferromagnetic super exchange interaction,  caused by the overlap of  
    dopant eigen functions. Then,   dopant spins are in a  random singlet state for dilute doping\cite{Bhatt82},   
     see Ref. \cite{aopreview} for a review.
    But there can be  
    other exchange mechanisms in dilute magnetic semiconductors,
    for example  the   Zener's double exchange coupling and  the p-d coupling, which are ferromagnetic. See Ref. \cite{dmsreview} for a review. 
    
    \subsection{Pseudo gap semimetal} 
    \label{pseudogap}

\begin{figure}[t!] 
   \includegraphics[width=0.27\textwidth]{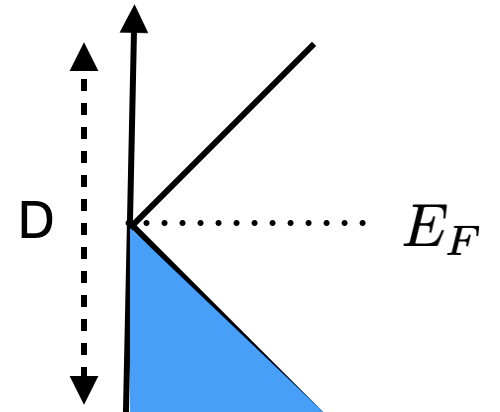}
  \includegraphics[width=0.27\textwidth]{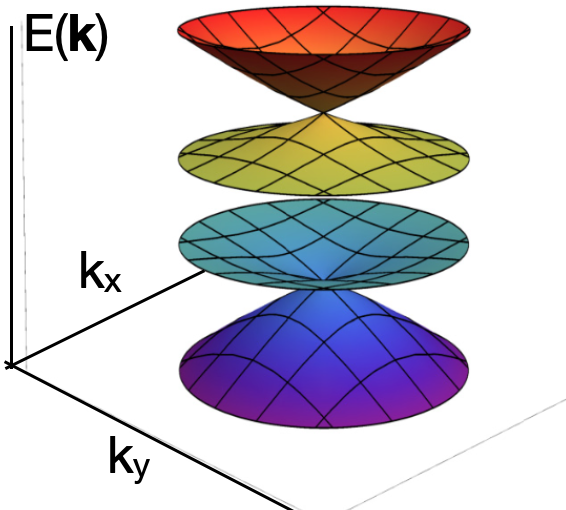}
    \includegraphics[width=0.44\textwidth]{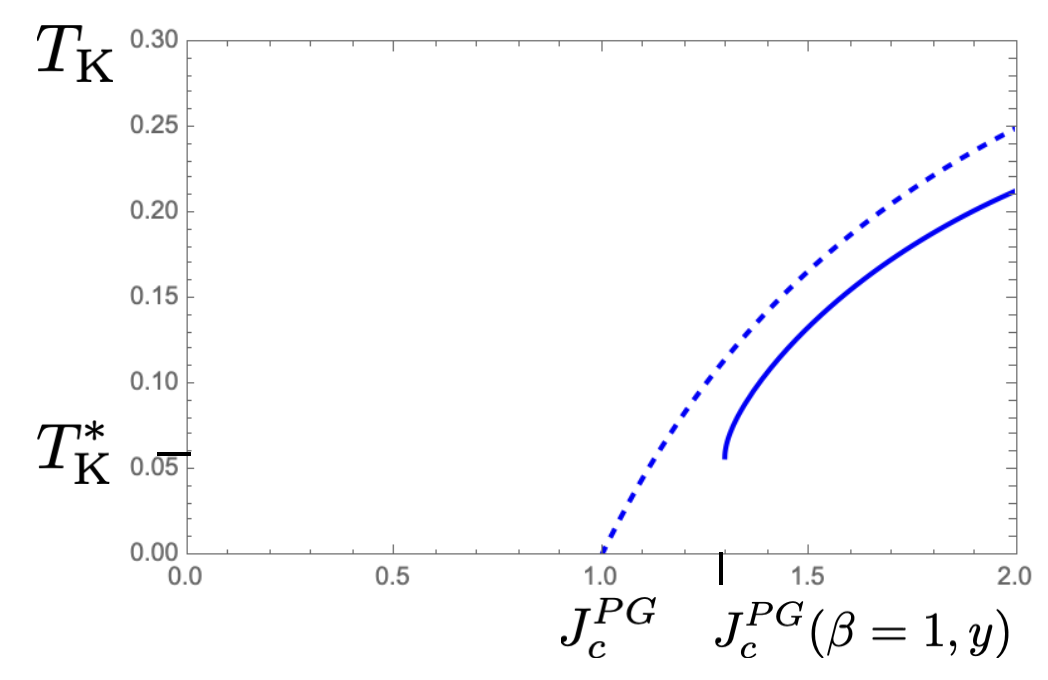}
 \caption{ Left: Schematic density of states with pseudo gap at Fermi energy $E_F$,  bandwidth $D$. Middle: 
 Quasiparticle dispersion  \index{quasiparticle dispersion}   of a  2D honeycomb  Kondo lattice.   Fig. taken from Ref. \cite{Lee2015}. Right: Kondo temperature in the presence of a pseudo gap with 
 power $\beta=1$ with RKKY coupling $y_1=1/(32 k)$  Eq. (\ref{tkpgy}) (full line) and without, Eq. (\ref{tkpg}) (dashed line). The Kondo temperature in the presence of RKKY coupling $y_1$ terminates at 
 the critical coupling  Eq. (\ref{jcpgy}) 
 $J_{c}^{PG}(\beta=1,y_1) >  J_{c}^{PG}(\beta=1)$,  Eq. (\ref{jcpg}) at the value $T_{\rm K}^*$,  Eq. (\ref{tkstar})  and jumps then to zero. }
 \label{qpspg}
\end{figure}

  A pseudo gap  \index{pseudo gap} 
    opens at the Fermi energy when the density of states vanishes as  $\rho (E)  \sim  |E- E_F |^{\beta}$  at the Fermi energy with power $\beta>0,$  as shown schematically in Fig.  \ref{qpspg} (left).  This occurs in numerous materials, like at the surface of topological insulators\cite{Hasan2010} and   in 
  graphene  
 \cite{CastroNeto2009}, where the electrons are confined to a   2-dimensional  honeycomb lattice. 
    The    low-energy excitations in graphene   are fermionic quasiparticles described by relativistic massless Dirac fermions. They are characterized by a linear dispersion relation,
    with two Dirac points, where the density of states vanishes linearly with energy, $\beta =1$. 
    Thus, 
   the question  arises,  what happens when these massless fermions are coupled to local magnetic moments. 
     The answer depends strongly on the magnitude of the exchange coupling $J$. 
  Plugging in the pseudo gap density of states 
   $\rho (E)   = \rho_0  |(E- E_F)2/D |^{\beta}$ 
  in the equation for the Kondo temperature, 
   Eq. (\ref{eq:FTKDOS}),     for $E_F = D/2$,   we find    for $J> J_c^{PG}(\beta)$
   with critical exchange coupling  \index{critical exchange coupling} 
   \begin{equation} \label{jcpg}
  J_c^{PG}(\beta)  = \beta/(2 N_0) =  \beta D/2,
    \end{equation}
      the Kondo temperature 
         \begin{equation} \label{tkpg}
       T_{\rm K} =  \frac{D}{2}  \left( 1 - \frac{J_c^{PG}(\beta)}{ J} \right)^{1/\beta}.
         \end{equation}
        Note that for $J \gg J_c^{PG}$, Eq. (\ref{tkpg})  converges to    $ T_{\rm K}^0  \sim  D/2 \exp ( -1/(2 N_0 J))$, the Kondo temperature in a metal.  
 There is 
      for $J <J_c^{PG}(\beta)$ no Kondo screening, as has been confirmed with non perturbative methods like the numerical renormalization group method in Refs. 
      \cite{pseudogap,ingersent,bulla}.  The concentration of free, unscreened magnetic moments
      at $T=0K$, 
       $n_{\rm FM} (J)$
       is a step function, $n_{\rm FM} (J) = n_{\rm M}$ for  $J <J_c^{PG}(\beta)$  and zero otherwise.

   It has been shown  in Ref. 
   \cite{Saremi2007}  with a large $N_K$-expansion that  there is a quantum phase transition  in the Kondo lattice on a  2-dimensional  honeycomb lattice   at at  critical coupling $J_c,$ even when neglecting the  RKKY coupling. Remarkably,  the energy dispersion of quasiparticles in 
    such a system with a pseudo gap has a direct gap, see Fig.   \ref{qpspg}  \cite{Lee2015}, where it was 
     shown that  the 
    Kondo-insulator gap is observable in the optical conductivity, in  stark contrast to the conventional  Kondo lattice system where the 
     Kondo-insulator gap is indirect, see Fig.  \ref{hf}.
     The  Dirac cones  become duplicated and shifted up and down in energy, respectively, as seen in Fig.   \ref{qpspg}.
      
      The 
      RKKY coupling  in presence of a pseudo gap at the Fermi energy
      is shorter ranged \index{RKKY coupling with pseudo gap} , 
      \begin{equation}
      J_{\rm RKKY} ({\bf R} )  =  \frac{ g({\bf R})  }{R^{d + \beta} },
      \end{equation}
       where $g({\bf R})$ is an oscillatory, non decaying function of ${\bf R}$ which may be anisotropic, 
       depending on the specific lattice. Here,  $R=|{\bf R}|$. 
    As an example,  in  a   2D honeycomb lattice like  graphene \index{graphene}, 
    where there  are  two sub-lattices $A$ and $B$, the RKKY coupling is 
    decaying with power $d+\beta =3$.
     The oscillatory function is 
    different, depending on whether  
     the magnetic moments are placed on the same sub-lattices, as given by\cite{Sherafati2011} 
     $g_{AA}({\bf R})  = - J^2 (1+ \cos ( \Delta {\bf K} \cdot { \bf  R} ) )$, 
     while on different sub-lattices, 
           $g_{AB} ({\bf R}) =  J^2  3 (1- \cos ( \Delta {\bf K}  \cdot { \bf R}  -2 \theta_R) ).$
       Here $ \Delta {\bf K} =   {\bf K}^+ -  {\bf K}^-,$ with    ${\bf K}^+ , {\bf K}^-$ 
 the reciprocal lattice   vectors  of  the two Dirac points. $ \theta_R$ is the 
    angle between the armchair direction and the position vector of the magnetic moment on the lattice. 

Thus, both  Kondo temperature and  RKKY coupling are diminished in the presence of a pseudo gap.
 This raises the 
 question how the pseudo gap modifies their competition. 
 As there is no Kondo screening for 
 $J < J_c^{PG}(\beta) = \beta D/2$, while the RKKY-coupling is finite for all $J$, 
  magnetic moments couple in that regime. To find out whether RKKY-coupling dominates the Kondo screening even 
 for larger exchange couplings 
 $J > J_c^{PG}(\beta)$, we need to solve the RG-equation, 
  Eq. (\ref{rgnm1}) with  RKKY interaction 
with a pseudo gap. 
    Insertion of   $\rho (E)   = \rho_0  |(E- E_F)2/D |^{\beta}$ into Eq. (\ref{rgnm1})
    and
     integration over $\Lambda$ up to the breakdown of perturbation theory
      at  scale $T_{\rm K}$, yields the  equation for the Kondo temperature  
     \begin{equation}
\label{tkypg}
\frac{J_c(\beta)}{J} - 1 +  \left( \frac{2 T_{K}}{D} \right)^{\beta}  + 
k \beta  y_{\beta}  J^{2}  N_0^2 \frac{D}{2 T_{K}} =0,
 \end{equation}
where $k = \ln (\sqrt{2} + 1)$, $ y_{\beta}$  the RKKY coupling 
function as modified  by a pseudo gap with power $\beta$,
and $J_c^{PG}(\beta)$ the critical coupling in absence of   $y_{\beta}$, Eq. (\ref{jcpg}). 
    
Let us consider as an example a pseudo gap with power $\beta =1$, 
as it  occurs in graphene and on the surface of topological insulators.
For  $J> J_c^{PG}(\beta=1, y)$ with  critical  exchange coupling  \index{critical exchange coupling} 
\begin{equation}
\label{jcpgy}
 J_c^{PG}(\beta=1, y_1)  = J_c^{PG} \frac{1- \sqrt{1-4 \sqrt{ k y_1}}  }{2 \sqrt{k y_1}},
 \end{equation} 
 where $J_c^{PG} = J_c^{PG}(\beta=1, y_1=0) = D/2,$ is the critical coupling when 
 neglecting the RKKY- coupling, 
we find the Kondo temperature 
\begin{equation}
\label{tkpgy}
  T_{K}  = \frac{D}{4} \left(1-\frac{J_c^{PG}}{J} + \sqrt{(1-\frac{J_c^{PG}}{J})^2 - k
  y_1 ( \frac{J}{J_c^{PG}})^2}\right).
 \end{equation}
 Note that at the critical coupling $ J_c^{PG}(\beta=1, y)$
 it  has a finite value, 
 \begin{equation}
 \label{tkstar}
 T_{K}^*  = \frac{D}{4} (1-\frac{J_c^{PG}}{J_c^{PG}(\beta=1, y_1)}),
 \end{equation}
 one half of the value it  has at that exchange coupling 
 $J_c^{PG}(\beta=1, y_1)> J_c^{PG}$, without   RKKY-coupling. 
  Thus, even in the presence of a pseudo gap, 
   the Kondo temperature jumps discontinuously to zero at $J_c^{PG}(\beta=1, y_1)$ as seen in Fig.  \ref{qpspg} (right),
   similar to what was found in a metal, Fig. \ref{dpdm}.
The critical coupling $J_c^{PG}(\beta=1, y_1)$ is an increasing function of the 
RKKY-coupling $y_1$ for $0< y_1 <1/(16 k),$ varying from $D/2$ to $D$.
For stronger coupling $y_1 > 1/(16 k)$  there is no finite  Kondo temperature and thus  no Kondo screening   for any coupling $J$. 

\section{  Spin competition in the presence of  disorder}  
\label{disorder}

   Any  real material has some degree of disorder. 
 In doped semiconductors it arises from the random positioning of the dopants themselves, in  heavy fermion materials 
 it may in addition  arise from structural defects
    or  impurities caused by atomic defects.
Disorder can cause Anderson localization \index{Anderson localization} , 
  trapping  electrons in the disorder potential.  Thus,  in order to  fully understand the physics of   electron systems with magnetic moments, 
we need to  understand how     Anderson localization affects the competition between  Kondo screening and RKKY coupling, and how that in turn affects Anderson localization. 
 Moreover,
   as noted  already early \cite{Anderson1977}, the physics of
      random systems is fully described only by
       probability distribution functions, not just averaged quantities.
             Thus, for electron  systems with randomly located  magnetic moments
             the derivation of physical properties
             requires the  knowledge  of  distribution functions of both 
             the Kondo temperature and the RKKY coupling\cite{Mott1976}, not just their averages.
        In fact, anomalous   distributions of the Kondo temperature $T_{\rm K}$  
        and the RKKY coupling can 
       give rise to non-Fermi-liquid behaviour, as measured for example  
      in  the low-temperature power-law divergence of the magnetic susceptibility
       in doped semiconductors close to the metal-insulator transition\cite{aopreview} .
        
       \subsection{Distribution of Kondo temperature and RKKY couplings}  
       
        Since the Kondo temperature  
 depends on the product of the local  exchange coupling $J$ and the 
density of states  at the Fermi energy
 $\rho$\cite{Kondo,Nagaoka65,Suhl65}, it is natural  to expect a distribution of the Kondo temperature, $P(T_K)$, when 
  $J$ and $\rho$ are distributed due to the random placement of the dopants,
  as has been pointed out  in Refs. \cite{Mott1976,Anderson1977,Miranda1996,Bhatt1992,Langenfeld1995,Cornaglia2006}. 
  
\begin{figure}[t!]
 \centering
   \includegraphics[width=0.3\textwidth]{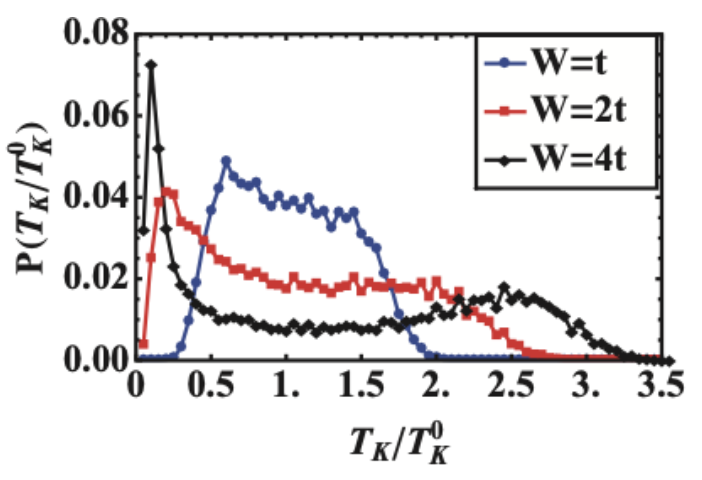}
  \includegraphics[width=0.28\textwidth]{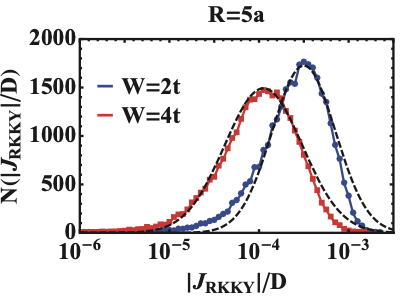}
  \includegraphics[width=0.27\textwidth]{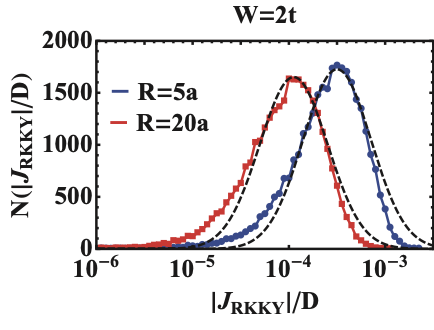}
 \caption{ Left: Distribution of Kondo temperature for  different disorder strengths $W$. 
 Middle: 
 distribution of  absolute value of the  RKKY coupling at fixed distance $R=5a$  for  different disorder strengths $W.$ 
 Right:  at fixed  $W=2t$ for   distances $R=5a,20a.$
 Dashed lines: 
 lognormal distribution with fitted parameters. 
All results are obtained   for 2D  tight binding model on square lattice, lattice spacing $a$, 
 with potential disorder, box distribution of  
 width $W$ in units of  hopping  amplitude $t$,  Fermi energy $E_F =t$.
    All Figs. taken from Ref. \cite{Lee2015}.}
 \label{distr}
\end{figure}

Indeed, 
  the  disorder  potential results in wave function amplitudes which vary 
  randomly, both
  spatially
  and with energy. 
           In a weakly disordered metal  different 
           wave functions  
           are correlated with each other in  
    a macroscopic energy interval of the order of  
     the elastic scattering rate $1/\tau$. This results 
     already in  weakly disordered metals in 
     a Kondo temperature distribution  \index{Kondo temperature distribution}  of  finite width in the thermodynamic limit\cite{micklitz06,prb2007}.     
     To model the disorder one adds 
         a disorder potential $V({\bf r})$ to the one particle Hamiltonian,
          which can be  assumed to be spatially uncorrelated and 
          white noise distributed with width given by 
          the elastic scattering rate
           $1/\tau.$  
      Using  the 1-loop equation for  the Kondo temperature 
      written in terms of the local density of states  \index{local density of states}  $ \rho (E, {\bf r} )$, 
   Eq. (\ref{eq:FTKDOS}), let us  
       rewrite it in terms of 
     the  disorder induced local deviations
         $ \delta \rho (E, {\bf r} )  =  \rho (E, {\bf r} ) - \langle  \rho (E, {\bf r} ) \rangle$,  where $\langle ... \rangle$ denotes the average over the disorder potential.  Denoting $T_K^{(0)}$ as  the Kondo temperature  
         obtained with  the 
         average local density of states, $\nu =  \langle  \rho (E, {\bf r} ) \rangle$ 
         we find the Kondo temperature for 
         a given realization of the disorder potential \index{disorder potential} \cite{Zarand96}
\begin{equation}
\label{eq:TK_transcd}
T_K = T_K^{(0)} \exp \left[ \int_{0}^{D} dE\, \frac{\delta
\rho({\bf r}, E)}{2 \nu (E-E_F)} \tanh \left( \frac{E-E_F}{2 T_K} \right)
\right].
\end{equation}
 Taking the square of the  logarithm of 
Eq. (\ref{eq:TK_transcd}) 
and performing the impurity average, 
we get
\begin{eqnarray} 
\label{lntk}
&&\left\langle \ln^2 \left( \frac{T_K}{T_K^{(0)}} \right) \right\rangle
 =\int_{0}^{D} dE\, \int_{0}^{D} dE' \nonumber \\ &
& \times \left\langle \tanh \left( \frac{E-E_F}{2 T_K^{(0)}} \right) \tanh
\left( \frac{E'-E_F}{2 T_K^{(0)}} \right) \frac{\delta \rho({\bf r},  E)}{2 \nu (E-E_F)} \frac{\delta
\rho({\bf r},  E')}{2 \nu (E'-E_F)} \right\rangle. 
\end{eqnarray}  
The 
 disorder  averaged  correlation function  \index{correlation function}  of  local density of states
 $ \langle \rho({\bf r},  E)  \rho({\bf r},  E') \rangle$ 
 is  governed at weak disorder $E_F \tau >1$
 by diffusion and Cooperon modes, as
  obtained  by summing up ladder diagrams, describing multiple 
  elastic scattering of the electrons  from  the impurity potential,
  for a review see Ref. \cite{mirlin}.
   Physically, these diffusion modes  \index{diffusion modes} account for the fact that electrons in a disorder potential do not move ballistically along  straight paths, but rather diffusively, such that the average square of the path length $\bf r (t)$
    on which an electron moves within  a time $t$ is given by 
    $\langle {\bf r (t)}^2 \rangle = D_e t,$ where $D_e = v_F^2 \tau/d$, is the diffusion constant. 
 Thereby one finds  for the standard deviation 
of the Kondo temperature  in the thermodynamic limit \cite{micklitz06,prb2007}
\begin{equation} 
\label{eq:tkstd}
\delta T_K   \approx T_K^{(0)}
  \left\{ \begin{array}{lr}  \frac{c_3}{(E_F \tau)
\sqrt{\beta}}\, \left[ \ln \left( \frac{1}{\tau\, T_K^{(0)}} \right)
\right] & {\rm in} ~d=3, \\ \frac{1}{\sqrt{3\pi E_F \tau
\beta}}\, \left[ \ln \left( \frac{1}{\tau\, T_K^{(0)}} \right)
\right]^{3/2} & {\rm in}  ~ d=2, \\
 2 \sqrt{ \frac{
\pi \sqrt{3}}{ k_F^2 A \beta} } (\tau T_K^{(0)})^{-1/4}
& {\rm in ~ quasi ~1-D~ wire ~ of ~ cross section ~} A, 
  \end{array} \right.
\end{equation}
   with $c_3$ a constant. 
 Note that  it  is larger with
time-reversal symmetry  \index{time reversal symmetry}  $\beta =1$ than when it is broken by 
 a magnetic field $\beta =2$, which
diminishes weak localization corrections. 
In the weak disorder limit,  the Kondo temperature has  a Gaussian distribution with width 
  given by Eq. (\ref{eq:tkstd}).
  However,  its distribution 
   becomes strongly  bimodal 
  as  disorder is increased further
   with an increasing weight at small Kondo temperature,
 see Fig.  \ref{distr} (left), where the distribution of
    $T_K$ is plotted,  obtained numerically
     for a   tight binding model on a square lattice 
 with potential disorder  of  box distribution and
 width $W$\cite{Lee2015}.
Furthermore, 
    a finite
    concentration  of {\it free} magnetic moments is found when electrons at the Fermi energy are localized. Since these effects can be explained by Anderson localization and multifractality, we will return to
   the  Kondo temperature  distribution after introducing these phenomena in the next chapters. 

As the RKKY coupling  is mediated by   conduction 
electrons,  it is strongly affected by   their  elastic scattering from the 
disorder, as well.  Indeed,  the disorder averaged RKKY coupling  \index{RKKY coupling} 
decays exponentially for  distances  larger than  the mean free path 
$l_e = v_F \tau$\cite{degennes}. This can be understood from the fact that 
 it depends on the product of the electron  wave function amplitudes at the locations ${\bf r}_i$ 
  and ${\bf r}_j$, see
Eq.
(\ref{eq:Jij}), and thereby on the electron phase difference between these two locations, 
 which 
the elastic scattering from disorder randomizes. However, 
 its geometrical average is hardly changed from its value in the clean limit\cite{zyuzin,Bulaevskii1986,Bergmann1987},
 which is valid even at stronger disorder, 
 as long as the distance between the magnetic moments is not larger than the 
 the localization length.
   The distribution of the RKKY coupling deviates from
     normal distribution already at weak disorder and
   converges to  a log-normal distribution  \index{lognormal distribution}  at stronger disorder,
   $P(x = \ln (|J_{RKKY}|/D) = \exp (- (x-x_0)^2/(2 \sigma^2))/(\sqrt{2 \pi} \sigma)$, with $x_0$ and $\sigma$ disorder dependent parameters, 
    as was derived with 
a field-theoretical approach \cite{Lerner1993}. That was  confirmed  numerically 
 for  2D disordered metals in Refs. 
     \cite{Lee2012a,Lee2012b,Lee2014},
    as seen in Fig.  \ref{distr}  (Middle and  Right). There, the distribution of
    $|J_{RKKY}|$ is plotted as obtained numerically
     for a   tight binding model on square lattice 
 with potential disorder and  box distribution of  
 width $W$ for various distances  between the magnetic moments 
 $R$ as compared to the lognormal distribution with fitted parameters (dashed lines)\cite{Lee2015}.
This  is  expected, since
      the amplitude of the RKKY coupling, Eq.
(\ref{eq:Jij}), is dominated by
      the local density of states at the Fermi energy, which  has at strong disorder at and close to  the Anderson localization transition a lognormal distribution\cite{mirlin}.
       The width $\sigma$ of the lognormal distribution 
       has been derived in Ref.  \cite{Lerner1993} to 
       scale with the elastic scattering rate  \index{elastic scattering rate}  as $1/\tau$ as 
       $\sigma \sim \tau^{-1/2}$, which has been confirmed numerically
        in 2D disordered systems \cite{Lee2014}, noting that 
$1/\tau = \pi W^2/(6D) $ with $D=8t$ the band width of the 2D tight binding model.

\subsection{Anderson localization - local spectral gaps}  

\begin{figure}[t!]
 \centering
   \includegraphics[width=0.26\textwidth]{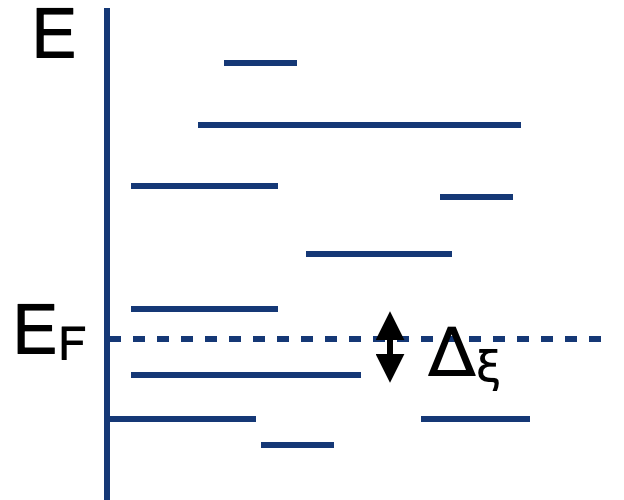}
   \includegraphics[width=0.25\textwidth]{ 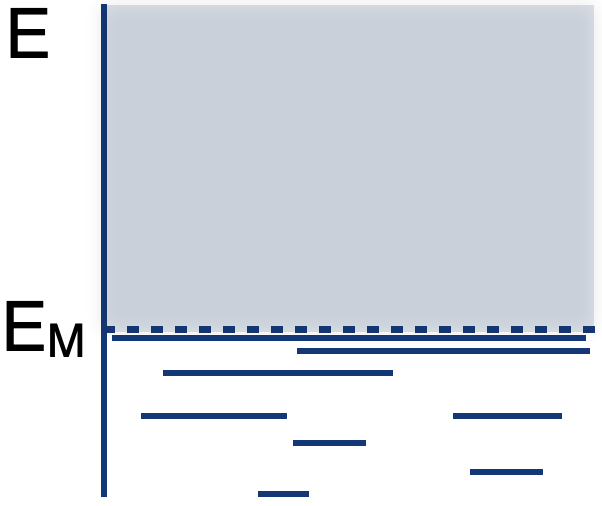}
 \caption{ Left: Spectrum of localized states with local level spacing $\Delta_{\xi}$.
 Right:  Spectrum with mobility edge $E_{\rm M}$, where for $E< E_{\rm M}$  all states
  are localized, while for $E>E_{\rm M}$ there is a continuum of extended states. At the mobility edge
    $E=E_{\rm M}$ there is a critical state.  }
 \label{loc}
\end{figure}

Disorder can result in 
Anderson localization  \index{Anderson localization} , where states are exponentially localized with  localization length $\xi$, 
forming a discrete spectrum with local level spacing $\Delta_{\xi}$ as sketched in Fig. \ref{loc} (left).
Since  electrons need then to  be thermally activated to 
contribute to a current, their resistivity is found  at  low temperature to  increase exponentially. The
interplay of   Anderson localization with spin correlations like  Kondo effect and RKKY interaction,  has  only recently
received increased attention, even hough  both  spin correlations
    and disorder effects are   relevant for many materials, including
   doped semiconductors close to the metal-insulator transition \cite{Loehneysen2011,aopreview},
   and typical heavy Fermion systems like  materials with
     4f or 5f atoms\cite{Miranda2005,Magalhaes2012}.
                         
 In $d=3$ and higher dimensions  mobility edges  \index{mobility edge} 
 $E_M$ appear 
  in the  band of eigen states of a Hamiltonian of 
  electrons moving in a disorder potential 
   $V({\bf r}).$  Then, the eigen states are found to be 
    localized with energy dependent localization length 
    $\xi(E)$
    at the band edges.
  There is a 
    delocalization transition  \index{delocalization transition}  at $E_M$ to a continuum of extended states,
     as sketched in Fig. \ref{loc} (right).
     where the locaization length diverges with a power law
     $\xi(E) \sim |E-E_M|^{-\nu}$ with critical exponent $\nu.$
For $d<2$ all states are localized for a Hamiltonian 
 of  noninteracting disordered electrons. 
 For $d=2$ all states are localized 
 in a disorder potential without magnetic field, or in weak magnetic field. 
  In the presence of a strong perpendicular magnetic field  in two dimensions,   critically extended states
   appear in the middle of  Landau bands,  which is known as  the integer quantum Hall transition. In  presence of spin-orbit interaction in two dimensions   there is a critical  delocalization transition.  
   Also, long range 
 interactions may cause a delolalization transition in two dimensions. For  reviews  on Anderson localization,  see Refs. \cite{Lee1985,kramermackinnon,Belitz1994,Efetov1997,KOK2005, Evers2008,Bhatt2021}.
 
 When eigen states are localized with localization length $\xi(E),$
   the spectrum is discrete with a local spacing 
   between the energy levels  $\Delta_{\xi} = 1/(\rho(E) \xi(E)^d),$ \index{local level spacing} 
   where $\rho(E)$ is the density of states at energy $E$ and $d$ is the dimension. 
   Thus, when placing magnetic moments in a disordered electron system, 
    and the Fermi energy $E_F$ is in the band of localized states, 
     the Kondo renormalization of exchange couplings stops 
     at energy scale $\Delta_{\xi}(E_F),$  since there are no states coupling to the magnetic moment 
     at lower energy.  Even though the gap is local,  as the exchange coupling 
      is local as well, this problem  is equivalent to the Kondo effect in the presence of a spectral gap, which we reviewed in section \ref{gap}.
      Thus, we  can conclude that there is 
       a critical exchange coupling $J_c^A$
        below which the magnetic moment remains unscreened, 
        where  \index{critical exchange coupling} 
       \begin{equation} \label{jcd}
          J^A_c(\Delta_{\xi}) = \frac{1}{2 N_0}
 \frac{1}{\ln (D/\Delta_{\xi} (E_F) )},         \end{equation}
 where $N_0 = N(E_F) = V_a \rho(E_F)/2$ is the density of states at the Fermi energy. 
    
    The RKKY coupling, on the other hand, is cut off   for 
    length scales exceeding the localization length at the Fermi energy 
    $R>\xi(E_F),$ but remains finite between magnetic moments whose distance
     is smaller,    $R < \xi(E_F).$
At and in the vicinity of  the mobility edge  $E_M,$               
     another phenomenon has to be taken into account, to understand the 
      competition between Kondo effect and RKKY-coupling there:
       the eigen function intensities have  multifractal distributions and 
        the intensities at different energy are power law correlated. 
         We  give a brief introduction to multifractality in the next chapter, 
          before reviewing its effects on spin correlations.

            \begin{figure}[t!]
 \centering
   \includegraphics[width=0.27\textwidth]{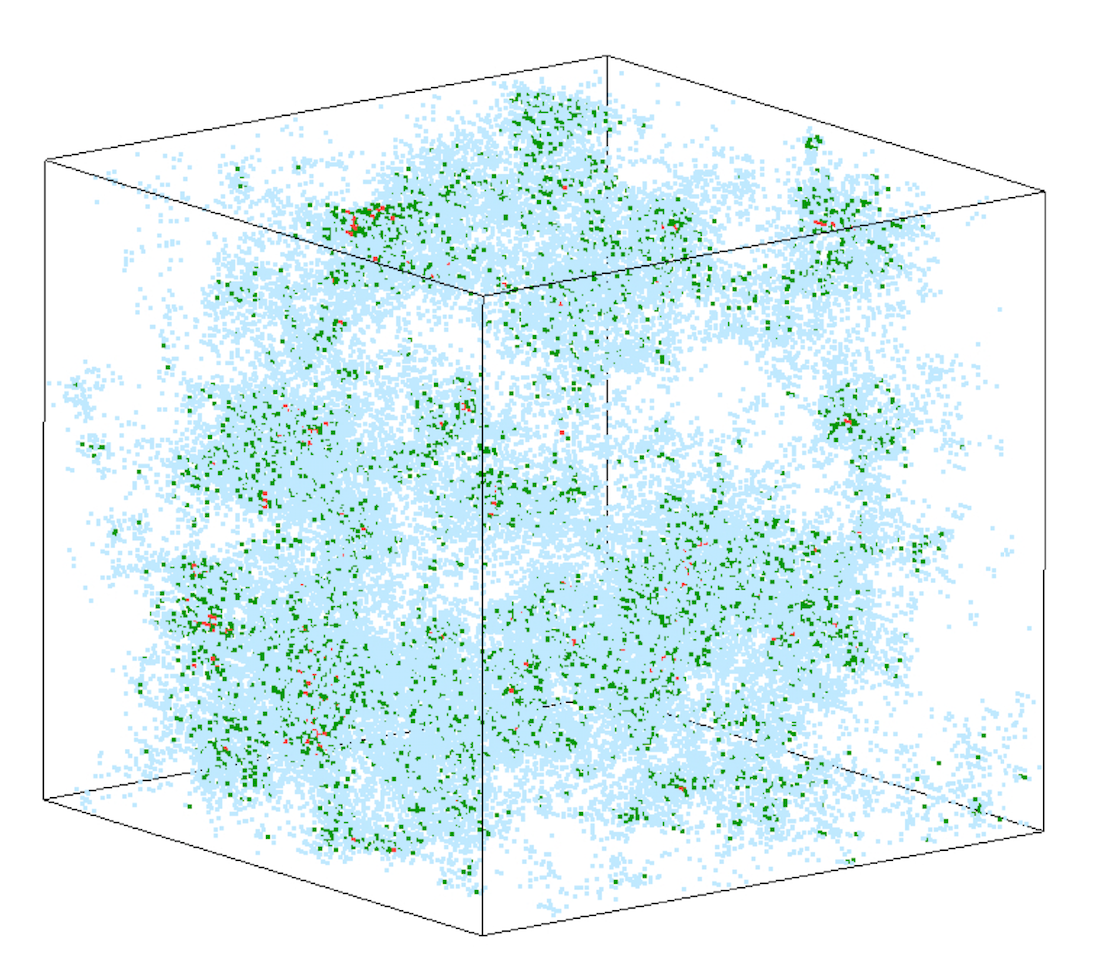}
   \includegraphics[width=0.65\textwidth]{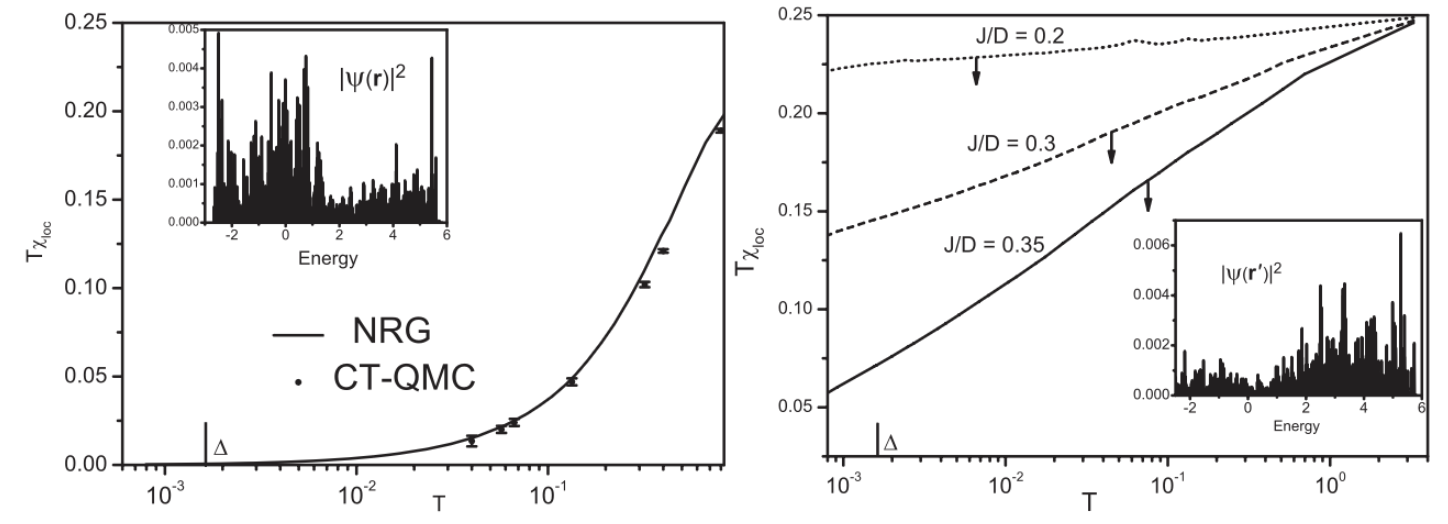}
 \caption{ Left:  
  Critical state  \index{critical  state}  intensity
   at $E=2t$ on  $d=3$ 
  Anderson tight binding model  
    ($N=10^6$ sites, disorder amplitude $W=16.5 t$), obtained by 
   exact diagonalization.  Colouring denotes  
   $ \alpha = -\ln |\psi|^2 / \ln L$ with
$\alpha \in [ 1.2,1.8]$ (red),  $ \alpha \in [1.8,2.4] $ (green), $\alpha \in [2.4,3.0]$(blue).
 Sites with higher intensity $\alpha < 1.2$ and lower intensity $\alpha > 3$ are not shown.
Thereby,   80 $\%$ of  total state intensity is visible.
   Fig. taken from Ref. \cite{Kats}.   Middle and Right:  Local spin susceptibility as function of temperature  $T$ for   spin $S=1/2$ coupled by exchange coupling  $J=0.35D,$
 disorder amplitude $W=2t$ in 2D lattice  ($L=70$), 
 obtained with numerical renormalization group  (lines) and continuous time quamtum Monte Carlo (dots) method,  at  sites  
where Kondo temperature $T_K$ is maximal (middle) and   where the  magnetic moment 
remains free (right). Arrows indicate $T_K^{(0)}(J)$. Insets: Intensity as  function of $E$
 ($E_F=0$). Fig. taken from Ref. \cite{zhuravlev}.  }
 \label{mf}
\end{figure}             

\subsection{Multifractality - local pseudo gaps}  
   
   Multifractality  \index{multifractality}  has been  observed
             experimentally in   disordered thin film 
            systems measuring   the 
       local density of states 
       by scanning tunnelling microscopy \cite{Richardella2010,Gao2020}.
In the vicinity of the Anderson delocalization transition  wave functions 
 have been shown to be 
strongly inhomogeneous,
             multifractal
          \cite{Evers2008}   
          and power law 
             correlated in energy\cite{powerlaw,cuevas}.   
              Since the  delocalization  transition is a 2nd order quantum phase transition,   
   the localization length  $\xi$ on the localized side of the transition, and the 
     correlation length  $\xi_c$  on the metallic side, diverge at  $E_M$ as  $\xi(E) \sim |E-E_M|^{-\nu}$ and 
    $ \xi_c(E) \sim |E-E_M|^{-\nu_c},$ where universality implies $\nu = \nu_c$.
     Thus,  at the  mobility edge there is  a critical state, 
  which is very sparse,  but spread over the whole sample, see  
 Fig. \ref{mf} (left),  where  the   intensity $|\psi_n({\bf r})|^2$  is plotted  for all sites of a 
 finite sample of  
 the 3D tight binding model with onsite disorder potential. 
The critical  eigen function intensities $|\psi_l({\bf r})|^2$ are found to 
scale with  linear size $L$ as,
\begin{equation}
\label{single}
P_q = L^d \langle\, |\psi_l({\bf r})|^{2 q} \rangle \sim L^{-d_{q} (q-1)}.
\end{equation}
   Critical states are  characterized by 
    multifractal dimensions $d_{q} <d,$  smaller than  spatial dimension $d$ 
    and changing with   power $q$.
The   local  intensity distribution 
of  a critical state is close to  a log-normal distribution function, as given by\cite{Evers2008}
\begin{equation}
\label{eq:Pone}
P(| \psi_l({\bf r}) |^2) \sim 
L^{\alpha_{\psi} - (\alpha_{\psi}-\alpha_0)^2/(2   \eta) },
\end{equation}
where $ \alpha_{\psi} = - \ln | \psi_l({\bf r}) |^2 / \ln L $ and $  \eta  = 2 (\alpha_{0}-d)$ with 
$\alpha_0>d$.   Then,  $d_{q}= d - q (\alpha_{0}-d) $
  for  $q< q_c$.  
  For $q>q_c= \alpha_{0}/\eta$
it  terminates at 
  $  \tau_{q_{c}}$\cite{Evers2008}. 
 Away from criticality 
          wave functions 
      show multifractality  at  
           length scales  smaller than  localization length $\xi(E)$ and  correlation length $\xi_c(E)$,
           respectively, where  moments scale with $\xi(E)$,  $\xi_c(E)$ 
           in multifractal dimensions $d_q$.   
   
            Another consequence of multifractality is that 
         intensities are power law correlated in energy\cite{powerlaw,cuevas} 
             within
      correlation energy $E_c \sim 1/\tau$ \index{correlation energy} .
 Given that  the intensity at
the critical energy $E_{k}=E_{M}$ is $|\psi_M ({\bf r})|^2 =
L^{-\alpha_{\psi}}$ the conditional
intensity of a state at energy $E_{l},$  is  relative to
the intensity of an extended state $L^{-d}$  given by  \cite{Kats} \index{conditional intensity} 
\begin{equation}
\label{cci}
I_{\alpha}  =  L^d \langle| \psi_{l} ({\bf r}) |^2
\rangle_{|\psi_M({\bf r})|^2=L^{-\alpha_{\psi}}}  \sim  \left|
\frac{E_{l} -E_{M} }{E_{c}} \right|^{\beta_{\alpha}},
\end{equation}
 for  $| E_{l} -E_{M} | < E_{c} \sim 1/\tau$. Thus 
 the intensity varies with a 
  power  law with power  $\beta_{\alpha}=(\alpha-\alpha_{0})/d$
  for   $| E_{l} -E_{M} | > \Delta$
 (when $E_{l}$ is closer 
to $E_{M}$  than the level spacing $\Delta,$
 the conditional
intensity 
reduces to the intensity itself,  $L^{-\alpha_{\psi}}$). 
At positions
 where the intensity 
is small,   $\alpha > \alpha_0,$ it
remains  suppressed within an energy range of order $1/\tau$ around $E_{M}$
  forming a {\it  local pseudogap } \index{local pseudogap}  with power $\beta_{\alpha}>0.$ 
  Indeed,  such local pseudogaps are found 
  numerically with only  small fluctuations, see the inset of 
  Fig. \ref{mf} (Right)
   for  a 2D disordered system with linear size $L<\xi$ at $E_F=0$.  
 When the
intensity  is larger than its typical value
$L^{-\alpha_{0}}$,  $\alpha < \alpha_{0}$,  it remains  enhanced within  an energy range of order $1/\tau$ around
$E_{M}$, increasing as a power law when $E_{l}$ approaches the
mobility edge. An example 
of such  strong enhancement at  $E_F =0,$
as obtained numerically for a 2D system with linear size $L<\xi,$ is shown in the inset of Fig. \ref{mf} (Middle).



\begin{figure}[t!]
 \centering
   \includegraphics[width=0.45\textwidth]{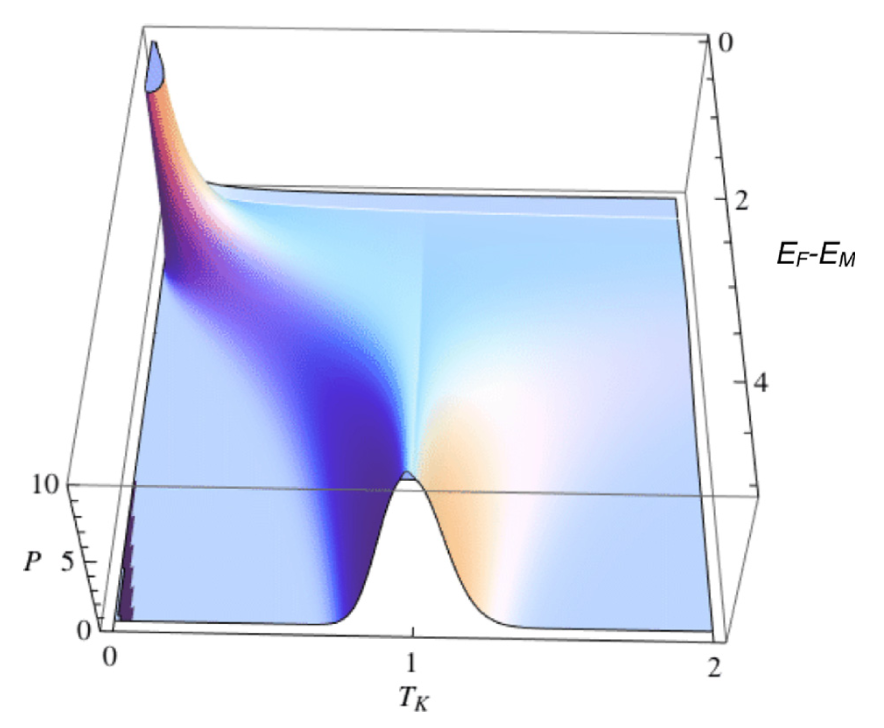}
     \includegraphics[width=0.5\textwidth]{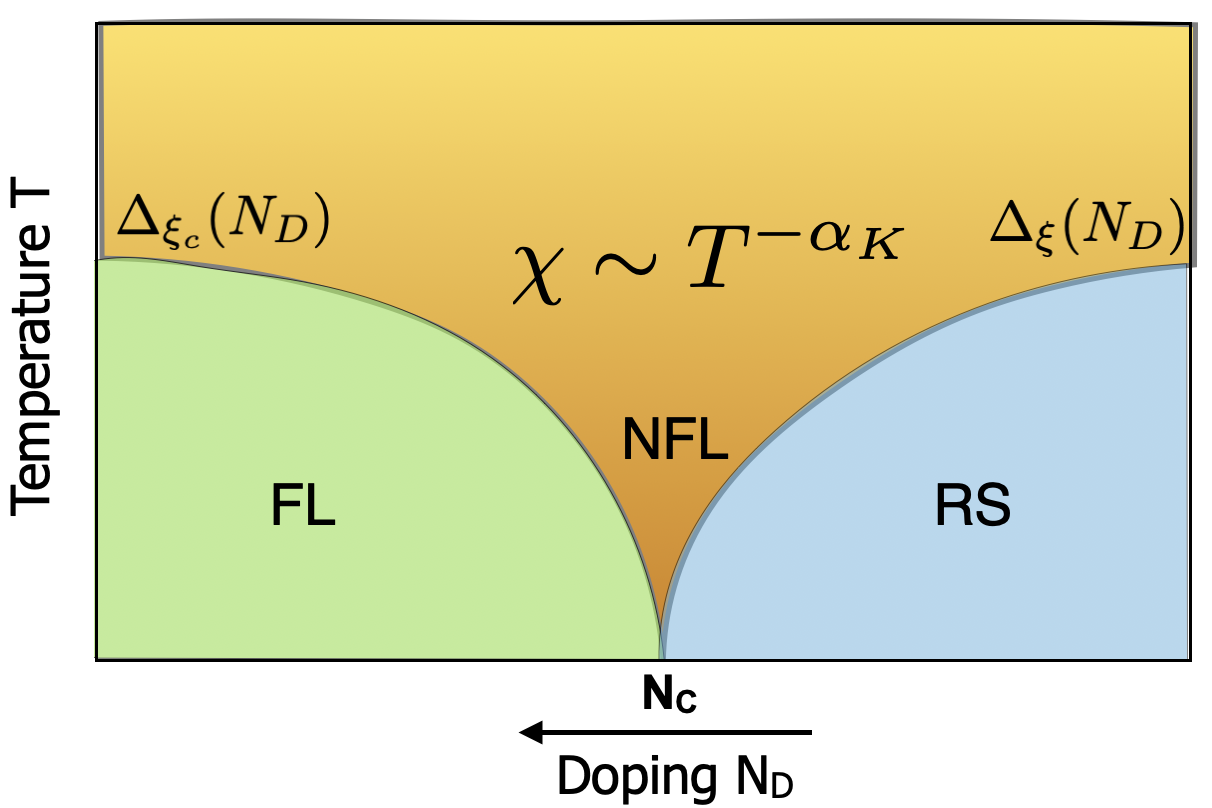}
 \caption{Left:  Kondo temperature $T_K-$distribution in units of $T_K^{(0)}$
 as function of distance to mobility edge $E_F - E_M$ in units of $E_c$ for exchange coupling 
 $J = D/5,$ derived analytically\cite{Kats}.  Fig. taken from Ref. \cite{Kats}.
 Right: Schematic phase diagram as function of doping concentration $N_D$, near  critical doping $N_c$.
 Non-Fermi liquid behaviour   at temperatures exceeding the scale $\Delta_{\xi_c}(N_D)$ on  the metallic side and $\Delta_{\xi}(N_D)$ 
 below which there is a Fermi liquid  (FL)
  due to Kondo screening, 
 and a random singlet state (RS), respectively.}
 \label{ptkc}
\end{figure}

  Implementing multifractality and power law correlation  \index{power law correlation}  of   intensities, 
             Kondo temperature $T_K$  is found by inserting  the 
            conditional intensity of state $l$,  Eq. (\ref{cci})  into Eq. (\ref{eq:FTK}),
\begin{equation}
\label{tkc3}
1 =  \frac{J \Delta
 }{2 D  E_{c}} \sum_{|\epsilon_{l} |< E_{c} }
 \left|
\frac{\epsilon_{l}  }{E_{c}} \right|^{\beta_{\alpha}-1}\tanh \left(\frac{\epsilon_l}{2 T_K } \right),
\end{equation} 
  where  the summation over $l$ is restricted  to energies within the energy interval 
   of  the correlation energy
   $E_{c} \sim 1/\tau$ around the mobility edge.  The power is given by 
$\beta_{\alpha} = (\alpha_{\psi}-\alpha_{0} )/d$
 for  $ \Delta <| E_{l} -E_{M} | < E_{c} $. 
Thus,  Eq. (\ref{tkc3}) defines the Kondo temperature in a system
with {\it pseudogaps} of power $\beta_{\alpha}$,  in the local density of states
   when it is
positive\cite{pseudogap} and the Kondo temperature is reduced at such sites.
On the other hand, at sites where
 power $\beta_{\alpha}$ is negative  $T_{K}$
is enhanced. 
          Eq. (\ref{tkc3})  can be solved analytically  yielding\cite{Kats}
\begin{equation}
\label{tkalpha}
\frac{T_K}{E_c} = \left[ \left(1 - 
  \frac{J_c^{PG}(\beta_{\alpha}  )}{J} \right) c_{\alpha} \right]^{\frac{1}{\beta_{\alpha}}},
\end{equation}
where $\beta_{\alpha} = (\alpha_{\psi} -
    \alpha_0)/d,$ 
    and the critical exchange coupling  \index{critical exchange coupling}  is 
     $J_c^{PG}(\beta_{\alpha}) = \beta_{\alpha} D/2$
 and $c_{\alpha} = (2 \alpha_{\psi}-\eta)/(\alpha_{\psi}-\eta/2+d).$. 
Thus, the Kondo temperature has the  form we found in the presence of a
pseudogap Eq. (\ref{tkpg}), with power $\beta_{\alpha}$.
For  $J<J_c^{PG}(\beta_{\alpha})$ the magnetic moment remains unscreened. 
Since 
$\alpha_{\psi} \in [0,\infty]$ is distributed, we find  that  $J_c^{PG}(\beta_{\alpha})$ and thereby 
$T_{K} (\alpha)$ are distributed, accordingly. 
 For the typical value
$\alpha_{\psi}=\alpha_{0}$ we recover $T_K$ of a clean system
$T_K(\alpha_{\psi} = \alpha_{0}) \sim E_{c} \exp(- 1D/(2J)) \sim T_K^{(0)}.$
            The derivation can be extended   into the vicinity of the mobility edge,
             where the (localization, correlation) lengths $(\xi, \xi_c)$ are finite, respectively
             by substituting in  $\alpha_{\psi}$  the system size $L$ by 
         $(\xi, \xi_c)$.
              Thereby, using a normal  distribution of  $\alpha_{\psi},$   the distribution of the Kondo temperature can be derived
               analytically, 
           as  plotted in  Fig. \ref{ptkc} (left) \cite{Kats}
           as function of energy distance to the mobility edge $E_M$.
              It evolves  
             from a Gaussian distribution with finite width in the weakly disordered metal regime to 
             a bimodal distribution with a divergent power law
               tail  \index{power law tail}  at the mobility edge. The enhanced weight at low Kondo temperatures                
              was shown 
              in Ref. \cite{Kats} to origin from 
               the opening of  the local pseudo gaps and is  given by 
               \begin{equation} \label{tail}
               P(0< T_K \ll T_K^0) \sim T_K^{-\alpha_K},
                \end{equation}
               with universal power $\alpha_K = 1- \eta/(2d),$  \index{universal power} 
            with 
         multifractal correlation  exponent  \index{multifractal correlation exponent}  $\eta= 2(\alpha_0-d)$.              
              The  magnetic susceptibility
           $ \chi (T) \sim n_{FM}(T)/T $ with  density of free moments $n_{FM} (T) = n_M \int_0^T d T_K  P( T_K)$,
               at temperature $T$ is found at the mobility edge as 
              \cite{Kats},
            \begin{equation}
            \chi (T) \sim  
   \left(\frac{T}{E_{c}} \right)^{-\alpha_K},
  \end{equation}
    diverging with a universal power law, \index{universal power} in good agreement with experimentally observed 
    Non-Fermi liquid behaviour  \index{Non-Fermi liquid}  in the magnetic susceptibility and specific heat \cite{Loehneysen2011}
      to which the  magnetic moments contribute 
  $C( T)  \sim T d n_{FM} (T)/dT$ as $C(T) \sim T^{\eta/(2d)}$. 
This result is also valid on the insulating side for  $T>\Delta_{\xi}$ and  on 
the metallic side  for $T>\Delta_{\xi_c}$, yielding the phase diagram
shown in Fig. \ref{ptkc} (right) with a  Non-Fermi liquid fan, caused by the 
distribution of Kondo temperatures due to the 
 multifractality in the vicinity of  the  mobility edge.  
    
 Numerical calculations of the   $T_K-$distribution  \index{Kondo temperature distribution} confirm
  the anomalous power \cite{Cornaglia2006,Slevin19} with a power which is very close to the analytical result Eq. (\ref{tail}).
   In Ref. \cite{zhuravlev}  the magnetic susceptibility was  obtained by  a full Wilson renormalization group (RG) calculation  \cite{Wilson75}   for a  2D disordered system
   finding anomalous power law divergence,  shown in Fig.  \ref{mf} at sites where the intensity is suppressed
   as in a local pseudogap. 
    Recently, it was pointed out that a more realistic  model of the Kondo impurity
    which takes into account anisotropies
    yields a modified distribution of  Kondo temperature  \cite{Debertolis2022}. It remains to be explored whether this will affect the low $T_K$ tail at the AMIT and thereby the anomalous  power
    of the magnetic susceptibility  $\alpha_K.$
    
       On the insulating side of the transition  there remains a finite density of magnetic moments 
       in the low temperature limit,
    since the Kondo screening becomes quenched by  Anderson localisation, 
   where renormalisation of the Kondo coupling becomes cutoff by the 
    local level spacing $\Delta_{\xi} = 1/(\rho \xi^3)$\cite{meraikh}. 
       Since these free moments   are still 
      weakly coupled to the electron system, they interact  with each other in the  vicinity of 
       the mobility edge  by 
    RKKY- like  exchange  interactions,   extending up to distances of the order of  the localization length $\xi$. 
    In the dilute doping 
                 limit the  indirect exchange interaction 
                becomes the superexchange interaction due to  the overlap of   hydrogen like impurity states, which 
is known to be antiferromagnetic.
 These  randomly positioned  magnetic moments have been modelled 
by a Heisenberg spin model with random antiferromagnetic
   short range, exponentially decaying exchange   interaction 
  \cite{Bhatt82}.
  In agreement with  experiments, numerical simulations and the implementation of  a
  cluster renromalisation group  no evidence for a  spin glass transition, 
  at which the magnetic susceptibility would peak and then decay to lower temperatures is found\cite{Bhatt82}.
  Rather,  the 
  magnetic susceptibility diverges at low temperature with a power law
  $\chi (T) \sim T^{-\alpha_J},$ with  $\alpha_J \le 1$\cite{Bhatt82}.
    In one dimension, the random antiferromagnetic short range Heisenberg model  is known to 
     flow at low temperature to the infinite randomness fixed point, where 
    the ground state  is formed of  randomly placed  spin singlets
     \cite{igloimonthus}.                              
   When the localisation length $\xi$ exceeds the Fermi  wave length, however,  
      the indirect exchange interaction oscillates in 
     sign with distance,   as the RKKY interaction
        in the metallic regime, but decays exponentially  beyond $\xi$.    
       In the next section we will  address the question, whether the  RKKY interaction or the Kondo effect wins 
         the spin competition in disordered electron systems in  the vicinity of the delocalization transition.

       \subsection{Doniach diagram of disordered systems}  
     
              \begin{figure}[h]
\begin{center}
\includegraphics[width=0.45\textwidth]{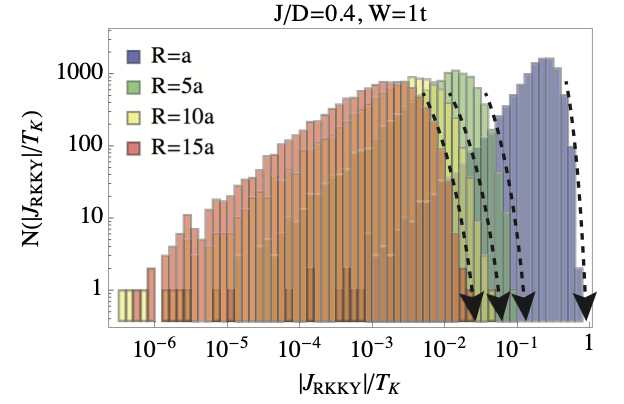}
\includegraphics[width=0.45\textwidth]{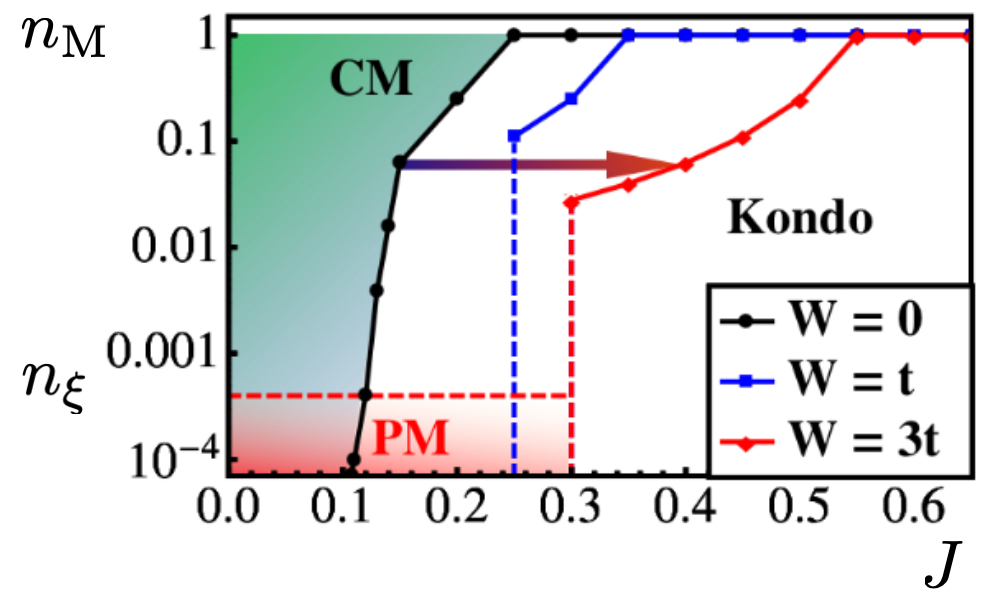}
\caption{ Left: Distribution of ratio  $|J_{RKKY}(R)|/T_K$
 for two magnetic moments at distance $R,$ placed randomly in  a 2D disordered lattice. Black  arrows: sharp cutoff. Right: Magnetic quantum phase diagram,
  critical density of magnetic moments
   $n_c$ as function of $J$  for various disorder strengths $W,$ as derived numerically from that cutoff, separating a coupled moment  \index{coupled moment}  (CM) from a Kondo screened phase.
     Horizontal dashed line separates the   free moment phase (PM) for $J < J_c (W)$,
     and $n_M < n_{\xi}$, where $n_{\xi}$ is the concentration, when there is no more than one magnetic moment within a localization length $\xi$. . Linear system size  $L = 100a$,  number of disorder configurations   $M = 30 000.$.
  Figs. taken  from 
 Ref.   \cite{Lee2014}.
}
\label{fig:ratio}
\end{center}
\end{figure}

           Extending the argument of  Doniach\cite{Doniach77}  \index{Doniach argument}  to  disordered systems
            where both  Kondo temperature $T_K$ and RKKY couplings are distributed,
            it is natural to study  next 
 the  distribution of the ratio  of both energy scales,  $J_{\textrm{RKKY}}({\bf r}_{i j})$
and $T_{K i}$. This  has been done in Ref.  \cite{Lee2014},
as shown  in  Fig. \ref{fig:ratio} (left) 
 for four distances $R$ between two magnetic moments, placed randomly in    a 2D disordered tight binding model. 
 While it is found to have a wide distribution for all $R$, there is a sharp cutoff, indicated by the black arrows.
  From   the condition that  $|J_{RKKY}(R)/T_K|<1$  
for all concentrations  $n_M = 1/R^d$  below  a critical value 
 $n_c= 1/R_c^d$, we can derive   $n_c$ accurately
   as function of exchange coupling  $J$. The resulting 
   quantum phase diagram is shown in  Fig. \ref{fig:ratio} (right) 
 for three disorder strengths $W.$ The Kondo dominated regime is 
  pushed to larger  $J$ as the disorder strength $W$ 
  is increased. 
  At strong disorder  a    phase  of uncoupled, paramagnetic local moments (PM)
  appears at small $n_M<n_{\xi}$, where $n_{\xi}$ is the concentration, when there is no more than one magnetic moment within the range of  a localization length $\xi$,
   as shown in  Fig. \ref{fig:ratio} (right) (horizontal red dashed line)\cite{Lee2014}. 
  
              \begin{figure}[h]
\begin{center}
\includegraphics[width=0.4\textwidth]{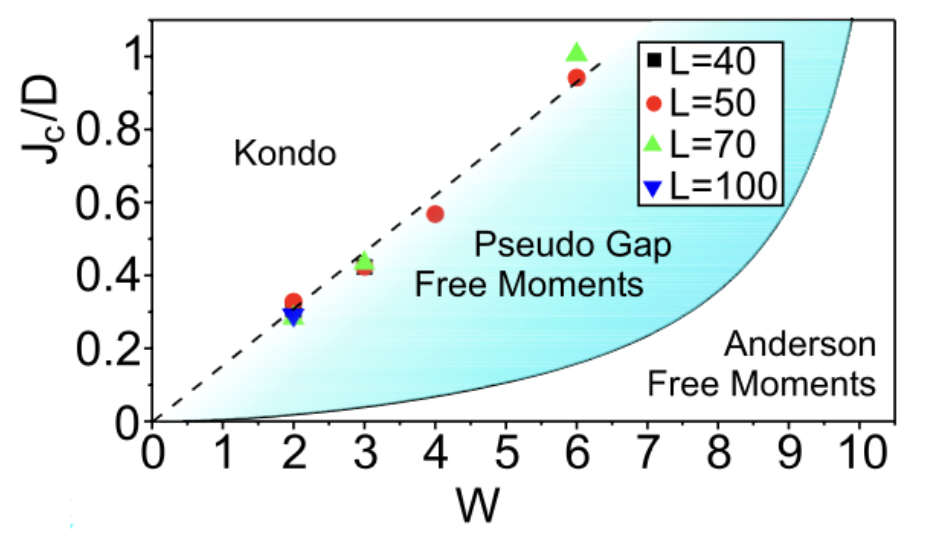}
\includegraphics[width=0.35\textwidth]{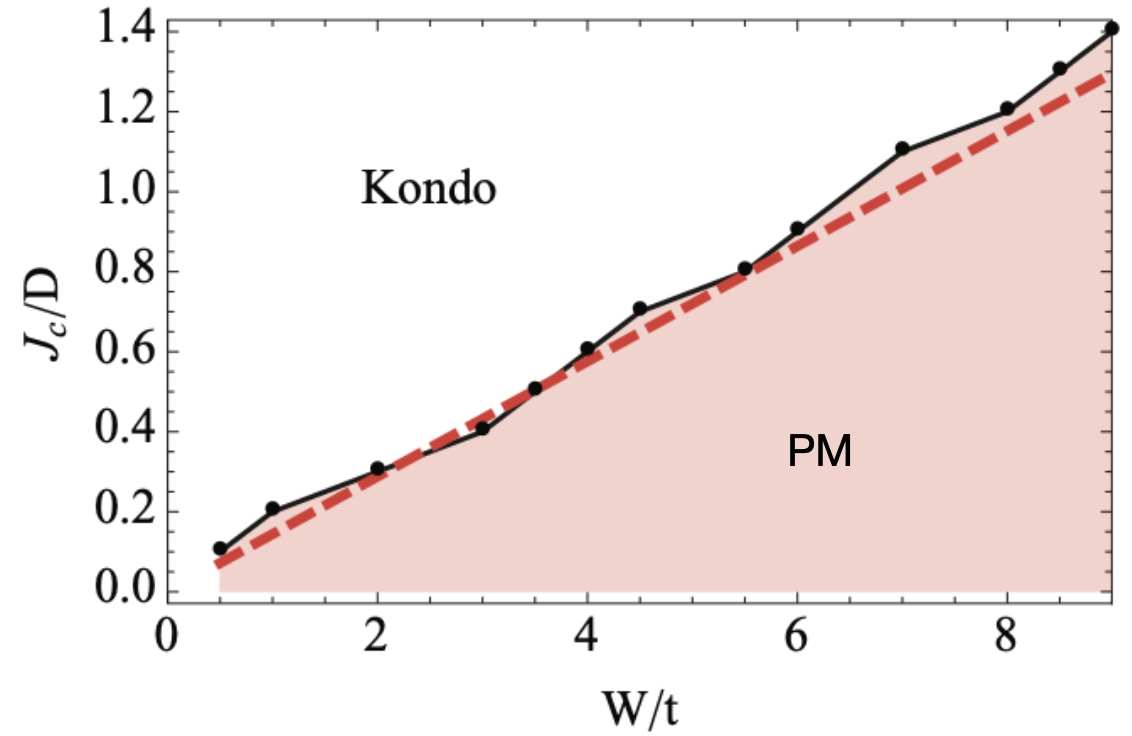}
\caption{ Left:  Critical coupling $J_c^{(1)}$
  as function of disorder amplitude $W,$ derived for  a 2D disordered
   tight binding model s function of
  disorder amplitude  $W$
  with      numerical exact diagonalization  for sizes $L$, 
  defined such that for   $J>J_c^{(1)},$
 there is no 
  unscreened magnetic moment in the sample. 
  Full line: $J_c^A(W)$, Eq.  (\ref{jcd}),
   with 2D localization length  $\xi = l_e  \exp(\pi E_F \tau),$ where $1/\tau = \pi W^2/(6D).$. Dashed line: guide to the eye. 
 Fig. taken  from 
 Ref.   \cite{zhuravlev}.
  Right: Same as left figure, but derived   
     by employing  Kernel polynomial method to Eq. (\ref{eq:FTK})
      for $L=100$\cite{Lee2014}. Dashed line: analytical
      function $J_c^{(1)} (W)$, Eq. (\ref{jc1}).
  Fig. taken  from 
 Ref.   \cite{Lee2014}.
}
\label{fig:fm}
\end{center}
\end{figure}

  In Fig. \ref{fig:fm} we show  the critical coupling $J_c^{(1)}$  \index{critical exchange coupling}  as function of disorder amplitude $W,$
  as defined to be the coupling   above which there remain no 
  unscreened magnetic moments in the sample.
  It  is derived for  a 2D disordered
   tight binding model s function of
  disorder amplitude  $W$
  with (left figure)
     numerical exact diagonalization  for several lengths $L$.
  The full line is a plot of  $J_c^A(W)$, Eq.  (\ref{jcd}),
   with 2D localization length  $\xi = l_e  \exp(\pi E_F \tau),$ where $1/\tau = \pi W^2/(6D).$. The   dashed line is a guide to the eye. 
    In Fig. \ref{fig:fm} (right)  we show results obtained 
     by employing  the Kernel polynomial method for   Eq. (\ref{eq:FTK})
      for system size $L=100$\cite{Lee2014}. The dashed line is the analytical
      function $J_c^{(1)} (W)$,  obtained from deriving  the  density of free moments due to local  pseudo gaps, yielding 
       $J_c^{(1)} =\sqrt{\eta} D/2,$ with 
       $\eta = 2(\alpha_0-d)/d$. In $d=2$, expansion in $1/g$, 
        $g= E_F \tau$, yields 
          $\eta = 2/(\pi g)$ and thus  
          \begin{equation} \label{jc1}
          J_c^{(1)} (W) =\sqrt{\frac{D}{3 E_F}} W. 
  \end{equation}
  The  good  agreement with numerical results supports the 
  result that the formation of free moments is 
   due to local pseudo gaps formed by multifractal intensity 
    distribution and correlations. 

\begin{figure}
\includegraphics[scale=0.5]{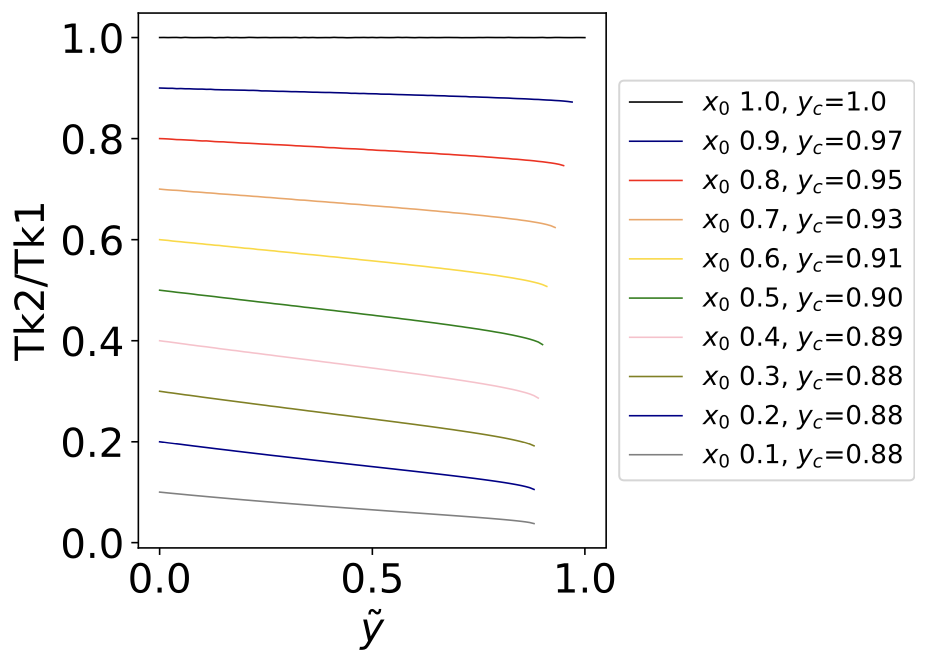}
\includegraphics[scale=0.5]{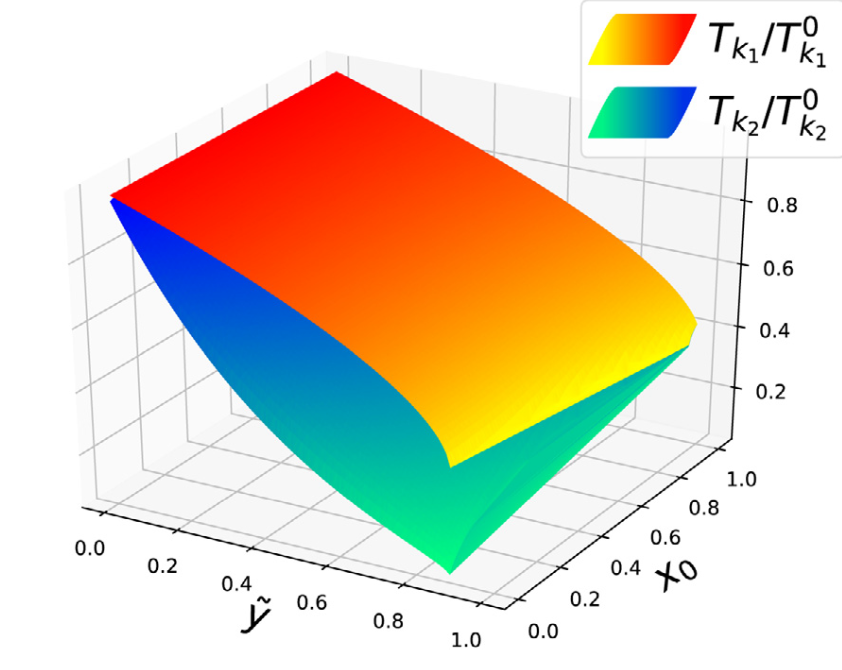}
  \caption{ Left:  The ratio of
  Kondo temperatures of  two magnetic moments
    $x=T_{K2}/T_{K1}$
    as function of dimensionless RKKY coupling parameter $\tilde{y}$,
    relative to its critical value $y_c$ for the homogenous system,
    for different bare Kondo temperature ratios
    $x_0= T^0_{K_{2}}/T^0_{K_{1}}$. It stops at a critical value $\tilde{y}_c (x_0)$, 
      relative to $y_c$ for the homogenous system.  Right: Kondo temperatures 
    $T_{K_1}$  and $T_{K_2}$  relative to their bare values as function of bare Kondo temperature ratios $x_0$ and the dimensionless RKKY coupling parameters $\tilde{y}.$
    Figs. are taken from Ref. \cite{Park2021}.
 }
 \label{fig:rkkykondo2}
\end{figure}

  To go beyond the Doniach argument for disordered systems let us  next 
   apply and extend  the self consistent approach of Ref. \cite{Nejati2017} \index{self consistent renormalization group} ,  reviewed in section 
 \ref{dd}.  Placing 
 two magnetic moments   at sites of a disordered electron system  with  
different  local density of states
    yields
  different bare Kondo temperatures $T^0_{K i} = D_0 \exp (-1/(2g^0_i))$, $i=1,2$\cite{Park2021}.
 By  solving 
  the   coupled RG-Eqs.  (\ref{rgnm1})  for two magnetic moments numerically,
  we find that both Kondo temperatures are reduced in  presence of 
 RKKY-coupling, see Fig.  \ref{fig:rkkykondo2}. 
  The initially smaller Kondo temperature $T_{K_{2}}$ is  suppressed more  
   than the  larger one  $T_{K_{1}}$.  Thereby,  their ratio $x = T_{K_{2}}/T_{K_{1}}$  decreases. Thus, we find that 
  the inequality  \index{inequality}  between  Kondo temperatures becomes enhanced   by RKKY coupling.
      Moreover,  the  quenching  of the Kondo screening by the RKKY coupling
 occurs already for smaller RKKY coupling, as seen in
 Fig.  \ref{fig:rkkykondo2} (left),
  the stronger
  the inhomogeneity and the smaller the ratio of the bare Kondo temperature  $x_0$. For the  smallest value,  $x_0=0.1$,   the breakdown  occurs
 at a critical value  
 $y_c(x_0 \ll 1)  = 0.88 y_c,$  where  $y_c$  is the one in the homogeneous system, Eq. (\ref{yc}).
The discontinuous jump of  both Kondo temperatures 
    $T_{K_1}$  and $T_{K_2}$  at $y_c(x_0)$  are seen in   Fig.  \ref{fig:rkkykondo2} (right),  plotted 
    relative to their bare values as function of bare Kondo temperature ratio $x_0$ and dimensionless RKKY coupling parameter $\tilde{y}.$
   Thus, we conclude that
disorder makes   Kondo screening    more easily quenchable by RKKY coupling.

For a finite density of  randomly distributed magnetic moments  $n_M,$  coupled by
random
local exchange couplings $J^0_i$ to  conduction electrons with  random  local density of states $\rho(E, {\bf r}_i ),$ one can extend this approach,
 solving the   coupled RG-Eqs.  (\ref{rgnm1}).
Every magnetic moment 
  has then, in general,  a different Kondo temperature, as they are placed at different positions  with different 
local couplings  $g_{0} (\bm{r}_i) =J^0_i  \rho(E, {\bf r}_i )$.
 As the RKKY coupling is itself  distributed widely in disordered systems
  \cite{Lerner1993,Lee2014} the long range
function $y(\bm{r}-\bm{r}')$ is distributed as well. We can thus 
 derive  the  distribution function of Kondo temperatures $T_{K}$ from  Eqs. (\ref{rgnm1}).
 We note  that the random distribution of RKKY-couplings is mainly
 due to the distribution of local couplings $g_{0} (\bm{r})$\cite{Lee2014}, while
 the function $y(\bm{r}-\bm{r}')$
  is only weakly affected  by the disorder.

Without the 
  RKKY-coupling   we found   at the mobility edge
 a bimodal distribution of $T_K$ with  one peak close to the
   Kondo temperature of the clean system and a
    power law divergent tail at low 
     $T_K$   \cite{Kats,Lee2014,Slevin19}, see Fig.  \ref{ptkc}.
  Since the RKKY-interaction  
   enhances inequalities between Kondo temperatures, we expect
    that it shifts 
   more weight to  the low Kondo temperature tail.
    This could  be checked quantitatively  by the solution of    Eqs. (\ref{rgnm1}),  but still needs to be done.
    
  Anderson localization is  weakened when time reversal and spin symmetry are broken by  magnetic scattering from magnetic moments
 \cite{Khmelnitskii1981,Wegner1986}.
 A finite magnetic  scattering rate 
$\tau_s^{-1}$  enhances the localization length     through the  parameter $X_s = \xi^2 /
L_s^2$,  where $L_s = \sqrt{D_e \tau_s}$ is the spin relaxation length \cite{Hikami1980} and 
 $D_e =  v_F^2\, \tau / 3$ the electron  diffusion coefficient.  When $X_s \ge
1$ the electron spin relaxes before it covers the area limited by  $\xi,$ 
and  Anderson localization is weakened. In 3D the mobility edge is thereby shifted towards stronger disorder by  magnetic scattering. 
 As Kondo screening of  magnetic moments diminishes  magnetic scattering, while  RKKY coupling tries to quench their  spins, resulting in magnetic scattering, the competition between these effects governs  Anderson localization and the position of the delocalization transition. 
  Treating the interplay between Anderson localization and Kondo screening, novel effects like a giant magneto resistance\cite{meraikh,Kats}, finite temperature delocalization 
  transitions and the  emergence of  a critical band\cite{Kats} have been 
  derived. 
 Experimentally,  the interplay of  Kondo effect, indirect exchange interaction 
   and Anderson localization has recently been 
    studied in a 2D experimental setup in a controlled way \cite{Zhang2022}. Thus, it 
    remains  an important 
    and experimentally relevant  problem, to develop  a
     self consistent treatment of   Anderson localization, Kondo screening and RKKY-coupling. 
     This problem has  been studied  
 with      the    disordered  Kondo lattice 
and  Anderson-Hubbard model with a variety of numerical methods. Each method comes with different  approximations,   providing different insights. These include  
mean field theories  of the Kondo lattice with an added RKKY coupling term  \cite{CastroNeto2000,kiselev,Magalhaes2012,Burdin2007,Tran2010,Tran2011,Burdin2021}, 
where fluctuations around the mean field theory have been studied with Ginzburg-Landau and nonlinear sigma model type actions. 
Statistical dynamical mean field theory based approaches \cite{Byczuk2009,Ulmke1995,Aguiar2006,Aguiar2009,Kotliar2003,Aguiar2013,Byczuk2005,Weh2021}, 
  Hatree-Fock based methods \cite{Milovanovic1989,Sachdev1998,Tusch1993}, quantum Monte Carlo method  \cite{Byczuk2011,Ulmke1997,Pezzoli2010,Costa2019}, 
   typical medium dynamics cluster approximations \cite{Jarrell2017,Jarrell2014}
 and cluster extensions of the  numerical renormalization group method \cite{MitchellLogan} have been applied.
 While we cannot review 
 all  results, some of them have been  reviewed in  Ref. \cite{Miranda2005},
 we would like to mention that 
            in Refs. \cite{Jarrell2017,Jarrell2014}
 the quasiparticle self energy  of the Anderson-Hubbard model  was
  derived as function of excitation energy $ \omega,$
     $Im \Sigma ( \omega) \sim \omega^{\alpha_{\Sigma}}.$
 It was 
   found to  behave  as a non-Fermi liquid   with power $\alpha_{\Sigma} (W) < 2,$
  which  becomes smaller with
      stronger   disorder amplitude $W$. 
     The non-Fermi liquid phase  extends down to  lowest energies
      at the mobility edge. Away from it, it is cutoff  by 
         $\Delta_{\xi} =1/(\rho \xi^d)$, and  $\Delta_{\xi_c} =1/(\rho \xi_c^d)$, respectively. 
         This is  in agreement with the phase diagram derived from the magnetic properties, reviewed above,  Fig. \ref{ptkc} (right). 
           The  typical medium dynamics cluster approximation
      employed in Ref.   \cite{Jarrell2017} 
       does not  include long range indirect exchange interactions.
       Thus,   the  study of the competition between   Anderson localization, Kondo screening and RKKY-coupling, remains 
a challenging problem.

\section{Conclusions and open problems}

\begin{figure}
\includegraphics[scale=0.5]{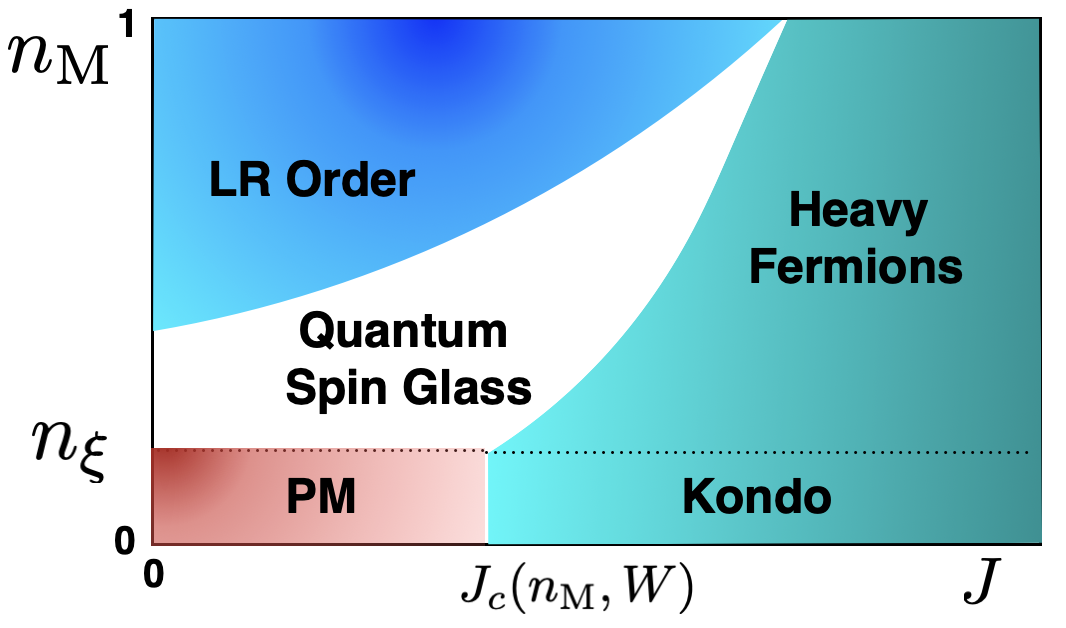}
  \caption{  Schematic quantum phase  diagram of Kondo lattice systems
  as function of magnetic moment density $n_{\rm M}$ and exchange coupling $J$
  for  fixed disorder  $W$.  The critical coupling $J_c(n_{\rm M},W)$
  separates spin coupled from Kondo screened phases.   Dark blue indicates higher transition temperatures to the long range ordered state (LR). Dark red indicates higher concentrations of free magnetic moments (PM), when electrons are localized and there is not more than one magnetic moment within the localization length $\xi$, $n_{\rm M}< n_{\xi} = \xi^{-d}$.  Dark petrol indicates higher Kondo and coherence 
   temperatures in 
   the dilute Kondo phase  and   heavy fermion phase.  }
 \label{fig:qpdKL}
\end{figure}

To conclude,  when 
 magnetic moments are immersed into the Fermi sea  of itinerant electrons,
  rich quantum physics emerges,  which is relevant for a 
a wide range of materials including  heavy Fermion systems,   materials with
     4f or 5f atoms, notably Ce, Yb, or U,  high temperature superconductors like 
     the cuprates, 
     but also 
     good metals with magnetic impurities, 
 doped semiconductors like Si:P close to the metal-insulator transition,
 2D materials like graphene and  topological insulators. While each material has its specific properties, requiring  detailed modelling, 
 their  electronic properties  are to some degree governed by the competition between 
  Kondo screening and indirect exchange couplings, which can be modelled
   by (disordered) Kondo lattices. 
We summarize the main results in the schematic  quantum phase diagram 
Fig.  \ref{fig:qpdKL}. 

While the  Doniach argument gives a good idea of the competition between Kondo screening and RKKY coupling,  a self consistent renormalization theory 
which  implements the RKKY coupling  into the Kondo renormalization 
yields a remarkably different result: 
 the Kondo temperature jumps  discontinuously to zero at a critical exchange coupling which is larger than expected with the Doniach argument. We reviewed the extension of  that framework  to a finite density of magnetic moments $n_{\rm M}$ to systems with a spectral (pseudo) gap  and to disordered systems, where the Kondo temperature is different for every magnetic moment. We have seen that disorder leads to a wide distribution of both Kondo temperature and RKKY couplings and tends to diminish the Kondo dominated phase, enhancing the critical coupling $J_c(n_{\rm M}, W)$. In the dilute limit we identified a paramagnetic phase, even at zero temperature, where the Kondo screening is prevented by local pseudo gaps, and the RKKY coupling between the dilute magnetic moments is cutoff by Anderson localization. In that regime, 
   the density of free magnetic moments is found to  decrease continuously with increasing $J$.
    There, an analytical formula for the increase of the critical coupling with disorder $W$, $J_c^{(1)}(n_{\rm M}, W)$ is available, Eq. (\ref{jc1}), which 
    was found   in 2D  to be in good agreement with numerical 
   results, see Fig. \ref{fig:fm}. 
   
As the density of magnetic moments is increased, there is  a 
  succession of  quantum phase transitions between     quantum  spin glass and ordered phases for couplings $J< J_c(n_{\rm M}, W)$, as
      shown schematically in Fig.  \ref{fig:qpdKL}. Since we cannot review here 
         the rich physics and variety  of these spin coupled phases, let us refer to the literature cited above and  
          recent reviews \cite{Coleman2007,Miranda2005,Magalhaes2012,aopreview}.
   For  $J> J_c(n_{\rm}, W)$,
      a transition  between a phase of dilute Kondo singlets and a heavy fermion 
      state is expected as $n_{\rm M}$ is increased. Experimentally, even for  dilute densities of about $n_{\rm M} = 0.05$, indications of 
   a   coherent Kondo lattice  were found in Si:P deep in the metallic phase \cite{Im2023} and  in the dilute Kondo lattice $CeIn_3$\cite{Park2024}. The 
     theory of the transition 
     from  dilute Kondo singlets to  heavy Fermions is still a challenging problem, as it requires to 
     solve  the dilute Kondo lattice  model, as   studied for example in \cite{Costa2019}. Taking fully  into account the disorder introduced by the random placements of magnetic moments is still a challenge. 
  Last but not least, even in the dense magnetic moment limit $n_{\rm M} \rightarrow 1$, 
    the mechanism for the emergence  of long range order at 
  the Kondo breakdown is under debate. Is it
due to the ordering of emerging  local moments, 
 or  a spin-density wave transition \cite{Coleman2007}  
  or a more complex mechanism, where ordered magnetic moments and  Kondo screening coexist in a spatially modulated state\cite{Li}?

\section{Acknowledgements} These lecture notes are dedicated to the memory of Peter Fulde. We  thank Hans Kroha and Keith Slevin for critical reading and feedback.



\clearchapter


\end{document}